\documentclass[apj]{emulateapj}

\usepackage{hyperref}
\usepackage{textcomp}
\usepackage{bm,color}
\usepackage{verbatim}
\usepackage{amsmath}
\usepackage{ulem}

\newcommand{\pb}{\textsc{Polarbear}}

\newcommand{\beq}{\begin{equation}}
\newcommand{\eeq}{\end{equation}}
\newcommand{\bea}{\begin{eqnarray}}
\newcommand{\eea}{\end{eqnarray}}

\renewcommand{\bf}{}

\begin{document}
\title{A Measurement of the Cosmic Microwave Background $B$-Mode Polarization Power Spectrum at Sub-Degree Scales from 2 years of POLARBEAR Data}


\author{The \pb\ Collaboration:
P.A.R. Ade\altaffilmark{32},
M. Aguilar\altaffilmark{6},
Y. Akiba\altaffilmark{31,19},
K. Arnold\altaffilmark{15},
C. Baccigalupi\altaffilmark{25,20},
D. Barron\altaffilmark{34},
D. Beck\altaffilmark{1},
F. Bianchini\altaffilmark{33},
D. Boettger\altaffilmark{24},
J. Borrill\altaffilmark{5,34},
S. Chapman\altaffilmark{12},
Y. Chinone\altaffilmark{14,26},
K. Crowley\altaffilmark{15},
A. Cukierman\altaffilmark{14},
M. Dobbs\altaffilmark{28},
A. Ducout\altaffilmark{26},
R. D\"unner\altaffilmark{24},
T. Elleflot\altaffilmark{15},
J. Errard\altaffilmark{1},
G. Fabbian\altaffilmark{21},
S.M. Feeney\altaffilmark{4,13},
C. Feng\altaffilmark{11},
T. Fujino\altaffilmark{35},
N. Galitzki\altaffilmark{15},
A. Gilbert\altaffilmark{28},
N. Goeckner-Wald\altaffilmark{14},
J. Groh\altaffilmark{14},
T. Hamada\altaffilmark{2,19},
G. Hall\altaffilmark{18},
N.W. Halverson\altaffilmark{3,16,8},
M. Hasegawa\altaffilmark{19,31},
M. Hazumi\altaffilmark{19,31,26,23},
C. Hill\altaffilmark{14},
L. Howe\altaffilmark{15},
Y. Inoue\altaffilmark{22,19},
G.C. Jaehnig\altaffilmark{3,16},
A.H. Jaffe\altaffilmark{13},
O. Jeong\altaffilmark{14},
D. Kaneko\altaffilmark{26},
N. Katayama\altaffilmark{26},
B. Keating\altaffilmark{15},
R. Keskitalo\altaffilmark{5,34},
T. Kisner\altaffilmark{5,34},
N. Krachmalnicoff\altaffilmark{25},
A. Kusaka\altaffilmark{29,17},
M. Le Jeune\altaffilmark{1},
A.T. Lee\altaffilmark{14,29,30},
E.M. Leitch\altaffilmark{7,27},
D. Leon\altaffilmark{15},
E. Linder\altaffilmark{34,29},
L. Lowry\altaffilmark{15},
F. Matsuda\altaffilmark{15},
T. Matsumura\altaffilmark{26},
Y. Minami\altaffilmark{19},
J. Montgomery\altaffilmark{28},
M. Navaroli\altaffilmark{15},
H. Nishino\altaffilmark{19},
H. Paar\altaffilmark{15},
J. Peloton\altaffilmark{10},
A. T. P. Pham\altaffilmark{33},
D. Poletti\altaffilmark{25},
G. Puglisi\altaffilmark{25},
C.L. Reichardt\altaffilmark{33},
P.L. Richards\altaffilmark{14},
C. Ross\altaffilmark{12},
Y. Segawa\altaffilmark{31,19},
B.D. Sherwin\altaffilmark{29},
M. Silva-Feaver\altaffilmark{15},
P. Siritanasak\altaffilmark{15},
N. Stebor\altaffilmark{15},
R. Stompor\altaffilmark{1},
A. Suzuki\altaffilmark{14,30},
O. Tajima\altaffilmark{19,31},
S. Takakura\altaffilmark{9,19},
S. Takatori\altaffilmark{31,19},
D. Tanabe\altaffilmark{31,19},
G.P. Teply\altaffilmark{15},
T. Tomaru\altaffilmark{19},
C. Tucker\altaffilmark{32},
N. Whitehorn\altaffilmark{14},
A. Zahn\altaffilmark{15}
}
 
\altaffiltext{1}{AstroParticule et Cosmologie, Univ Paris Diderot, CNRS/IN2P3, CEA/Irfu, Obs de Paris, Sorbonne Paris Cit\'e, France}
\altaffiltext{2}{Astronomical Institute, Graduate School of Science, Tohoku University, Sendai, 980-8578, Japan}
\altaffiltext{3}{Center for Astrophysics and Space Astronomy, University of Colorado, Boulder, CO 80309, USA}
\altaffiltext{4}{Center for Computational Astrophysics, Flatiron Institute, 162 5th Avenue, New York, NY 10010, USA}
\altaffiltext{5}{Computational Cosmology Center, Lawrence Berkeley National Laboratory, Berkeley, CA 94720, USA}
\altaffiltext{6}{Departamento de F\'isica, FCFM, Universidad de Chile, Blanco Encalada 2008, Santiago, Chile}
\altaffiltext{7}{Department of Astronomy and Astrophysics, University of Chicago, Chicago, IL 60637, USA}
\altaffiltext{8}{Department of Astrophysical and Planetary Sciences, University of Colorado, Boulder, CO 80309, USA}
\altaffiltext{9}{Department of Earth and Space Science, Osaka University, Toyonaka, Osaka 560-0043, Japan}
\altaffiltext{10}{Department of Physics \& Astronomy, University of Sussex, Brighton BN1 9QH, UK}
\altaffiltext{11}{Department of Physics and Astronomy, University of California, Irvine, CA 92697-4575, USA}
\altaffiltext{12}{Department of Physics and Atmospheric Science, Dalhousie University, Halifax, NS, B3H 4R2, Canada}
\altaffiltext{13}{Department of Physics, Imperial College London, London SW7 2AZ, United Kingdom}
\altaffiltext{14}{Department of Physics, University of California, Berkeley, CA 94720, USA}
\altaffiltext{15}{Department of Physics, University of California, San Diego, CA 92093-0424, USA}
\altaffiltext{16}{Department of Physics, University of Colorado, Boulder, CO 80309, USA}
\altaffiltext{17}{Department of Physics, University of Tokyo, Tokyo 113-0033, Japan}
\altaffiltext{18}{Harvard-Smithsonian Center for Astrophysics, Harvard University, Cambridge, MA 02138, USA}
\altaffiltext{19}{High Energy Accelerator Research Organization (KEK), Tsukuba, Ibaraki 305-0801, Japan}
\altaffiltext{20}{INFN, Sezione di Trieste, Via Valerio 2, 34127 Trieste, Italy}
\altaffiltext{21}{Institut d'Astrophysique Spatiale, CNRS (UMR 8617), Univ. Paris-Sud, Universit\'e Paris-Saclay, b\^at. 121, 91405 Orsay, France}
\altaffiltext{22}{Institute of Physics, Academia Sinica, 128, Sec.2, Academia Road, Nankang, Taiwan}
\altaffiltext{23}{Institute of Space and Astronautical Science (ISAS), Japan Aerospace Exploration Agency (JAXA), Sagamihara, Kanagawa 252-0222, Japan}
\altaffiltext{24}{Instituto de Astrof\'isica and Centro de Astro-Ingenier\'ia, Facultad de F\'isica, Pontificia Universidad Cat\'olica de Chile, Av. Vicu\~na Mackenna 4860, 7820436 Macul, Santiago, Chile}
\altaffiltext{25}{International School for Advanced Studies (SISSA), Via Bonomea 265, 34136, Trieste, Italy}
\altaffiltext{26}{Kavli IPMU (WPI), UTIAS, The University of Tokyo, Kashiwa, Chiba 277-8583, Japan}
\altaffiltext{27}{Kavli Institute for Cosmological Physics, University of Chicago, Chicago, IL 60637, USA}
\altaffiltext{28}{Physics Department, McGill University, Montreal, QC H3A 0G4, Canada}
\altaffiltext{29}{Physics Division, Lawrence Berkeley National Laboratory, Berkeley, CA 94720, USA}
\altaffiltext{30}{Radio Astronomy Laboratory, University of California, Berkeley, CA 94720, USA}
\altaffiltext{31}{SOKENDAI (The Graduate University for Advanced Studies), Hayama, Miura District, Kanagawa 240-0115, Japan}
\altaffiltext{32}{School of Physics and Astronomy, Cardiff University, Cardiff CF10 3XQ, United Kingdom}
\altaffiltext{33}{School of Physics, University of Melbourne, Parkville, VIC 3010, Australia}
\altaffiltext{34}{Space Sciences Laboratory, University of California, Berkeley, CA 94720, USA}
\altaffiltext{35}{Yokohama National University, Yokohama, Kanagawa 240-8501, Japan}

\begin{abstract}
We report an improved measurement of the cosmic microwave background~(CMB) $B$-mode polarization power spectrum with the \textsc{Polarbear}\ experiment {\bf at 150~GHz}. By adding new data collected during the second season of observations (2013--2014) to re-analyzed data from the first season (2012--2013), we have reduced twofold the band-power uncertainties. The band powers are reported over angular multipoles $500 \leq \ell \leq 2100$, where the dominant $B$-mode signal is expected to be due to the gravitational lensing of $E$-modes. We reject the null hypothesis of no $B$-mode polarization at a confidence of 3.1$\sigma$\ including both statistical and systematic uncertainties. We test the consistency of the measured $B$-modes with the $\Lambda$ Cold Dark Matter ($\Lambda$CDM) framework by fitting for a single lensing amplitude parameter $A_L$\ relative to the Planck~2015\ best-fit model prediction. We obtain $A_L = 0.60 ^{+0.26} _{-0.24} ({\rm stat}) ^{+0.00} _{-0.04}({\rm inst}) \pm 0.14 ({\rm foreground}) \pm 0.04 ({\rm multi})$, where $A_{L}=1$ is the fiducial $\Lambda$CDM\ value.
\end{abstract}


\thanks{
	\footnotesize{
		Corresponding authors: Yuji~Chinone, Julien~Peloton.
		\\ \hspace*{0.2in}\tt{chinoney@berkeley.edu, j.peloton@sussex.ac.uk}
	}
}

\section{Introduction}
\setcounter{footnote}{0}
The polarization of the cosmic microwave background~(CMB) encodes broad cosmological information that is the focus of current and future generations of CMB experiments.
The pattern of linear polarization separates into gradient-like $E$-mode and curl-like $B$-mode components.
{\bf On one hand, $E$-mode polarization is generated by the same scalar density fluctuations in the ionized plasma before recombination that generate temperature anisotropies.
In contrast, $B$-mode polarization is not generated by these scalar perturbations,
and could be generated by either tensor perturbations~(gravitational waves) from inflation or by conversion of $E$-modes to $B$-modes by gravitational lensing along the line of sight.}
On degree scales, where the inflation scenario predicts $B$-mode polarization of the CMB from primordial gravitational waves, no {\bf such} signal has yet been detected~\citep{bkp,planck2015XX,bkVI}.
In addition, gravitational lensing, which is not related to this primordial signal, induces a characteristic peak in the BB angular power spectrum at $\ell \sim 1000$. In this case, the {\bf primary} $E$-mode spectrum {\bf is} converted to $B$-modes by lensing from the large-scale distribution of matter.
{\bf Several experiments have begun to measure this lensing $B$-mode signal,
including \textsc{Polarbear}~\citep{pb2014c}, \textsc{bicep}2~\citep{bkI}, \textit{Keck Array}~\citep{bkV}, SPTpol~\citep{Keisler2015},
cross-spectra between \textsc{bicep}2/\textit{Keck Array}\ and Planck~2015\ $B$-mode maps~\citep{bkp,bkVI}, and \textsc{actpol}~\citep{2016arXiv161002360L}.}

This paper reports results from the \textsc{Polarbear}\ telescope, which has a $3\farcm 5$ resolution, giving its best sensitivity through intermediate angular scales around the $\ell \sim 1000$ lensing peak.
A previous \textsc{Polarbear}\ publication~\citep{pb2014c}, hereafter PB14, reported the first direct measurement of a non-zero $B$-mode signal, with modest significance.
In this paper, we present an improved measurement of the angular power spectrum as measured by \textsc{Polarbear}\ in the same survey areas~{\bf (a few degrees across for each sky patch)}, to greater depth.
The procedure used to analyze the dataset shares many similarities with that described in PB14. In particular, the analysis is conducted blindly, with angular power spectra revealed only when the data pass a series of null tests and systematic error checks. 
We also incorporate many new features and improvements in the analysis as well as an improved discussion of the contribution of astrophysical foregrounds. 
We include new data taken between {\bf October}, 2013 and April, 2014 (hereafter second season), but also a re-processing of the first season data~(May, 2012 -- June, 2013) in the light of the changes, which expands the total data volume with respect to PB14 by a factor of 61\%.
{\bf This means 62\% of the total data is reprocessed and 38\% of the total data is newly processed.}
We therefore refer the reader to descriptions in PB14 when necessary, emphasizing here only the additional steps. 

In Section~2 we describe the two seasons of data collected by the \textsc{Polarbear}\ instrument. 
In Section~3 we discuss the calibration procedure, and the data analysis steps are outlined in Sections~4, 5 and 6. 
In Section~7 we show the power spectra results, and in Section~8 we draw our conclusions.
 
\section{First and second season observations of the \textsc{Polarbear}\ instrument}
\label{sec:data}

\textsc{Polarbear}{} is a CMB experiment that has been observing from the 2.5\,m Huan Tran Telescope since January 2012. 
The telescope is located at the James Ax Observatory at an elevation of 5{,}190~m in the Atacama Desert in Chile. 
The \textsc{Polarbear}{}  receiver consists of an array of 1{,}274{} transition edge sensor~(TES) bolometers cooled to 0.3~K and observing the sky through lenslet-coupled double-slot dipole antennas. Within the array, the bolometers are grouped into 7 different wafers.
More details on the receiver and telescope can be found in \citet{Arnold_SPIE2012} and~\citet{Kermish_SPIE2012}. 
A cold half-wave plate (HWP) is positioned on the sky side of cryogenic lenses in the receiver. While this HWP was stepped almost daily for the first half of the first season, it was not stepped during the rest of the first season or during the second season. We discuss the role of the HWP on the mitigation of instrumental systematic effects in Sec.~\ref{sec:systematic_pipeline}.
 
The \textsc{Polarbear}{} observing strategy is described in PB14, and is summarized here.
We observe three CMB fields, with each one visible for 6 to 8 hours per day. 
The three patches are centered at (RA, Dec)=({4$^\mathrm{h}$40$^\mathrm{m}$12$^\mathrm{s}$}, {$-45$$^\circ$00\arcmin}), ({11$^\mathrm{h}$53$^\mathrm{m}$0$^\mathrm{s}$}, {$-0$$^\circ$30\arcmin}),
and ({23$^\mathrm{h}$1$^\mathrm{m}$48$^\mathrm{s}$}, {$-32$$^\circ$48\arcmin}) which we call RA4.5, RA12 and RA23, respectively.
We divide the observations into constant elevation scans~(CESs) during which the telescope scans back and forth in azimuth at constant elevation. 
We call each sweep in azimuth a subscan.
After approximately 15~minutes, when the patch has moved out of the field of view, we adjust the elevation and begin another CES.

The total observation time for the two seasons and the three CMB patches is 4{,}700~hours, corresponding to 33\% of the total calendar time available for the two seasons.
Of this time, 2{,}800~hours pass all the data quality checks described in Sec.~\ref{sec:data-selection} and are used to compute the power spectrum. 
As mentioned, this is an increase in data volume of 61\% over PB14. 

The process of extracting cosmological results from raw time-ordered data~(TOD) can be summarized in three steps:
calibration of the raw data, map-making, and power-spectrum estimation.
In the following two sections, we describe these processes.
 
\section{Calibration}
\label{sec:calibration}{\bf Before creating maps of CMB polarization anisotropy, we characterize detector and telescope performance.}
This includes reconstructing the telescope pointing, measuring the beam, calibrating the detector time-ordered data (TOD), and determining the detectors' polarization properties. 
We will describe each of these four steps in the following sections.

\subsection{Pointing}

To go from TOD to maps, we need to know where each detector was pointing as a function of time. 
We determine the pointing model using observations of bright extended and point-like millimeter sources selected from known source catalogs~\citep{vlanorth, at20g} across a wide range in azimuth and elevation. 
The approach is similar to that of PB14, although the model had to be extended to handle the increased pointing data volume and sky coverage. 
Relative to PB14, we use three times more pointing data across a $\sim19\%$ ($\sim30\%$) larger range in azimuth (elevation). 
Five new parameters were added to the pointing model to enable the pointing to be reconstructed over a larger fraction of the sky. 
The first was a timing error, i.e., an offset in hour angle per period of time, caused by small synchronization errors between clocks and ephemerides. 
The next two, solar radiation flexure in azimuth and elevation, account for the small deformation of the telescope due to temperature gradients related to the Sun's position. 
The final two, ambient temperature flexure in azimuth and elevation, are similar except with the deformation correlated with the ambient temperature \citep{fmatsuda2017}. 
Neglecting the four new flexure terms would worsen the RMS pointing uncertainty by 8\arcsec.

The final pointing accuracy is similar to PB14, with an RMS scatter measured on known radio source positions of 27\arcsec\ and 30\arcsec\ for seasons 1 and 2 respectively.

\subsection{Beam} \label{sec:beam}

\begin{figure}[!htbp]
\includegraphics[width=.45\textwidth]{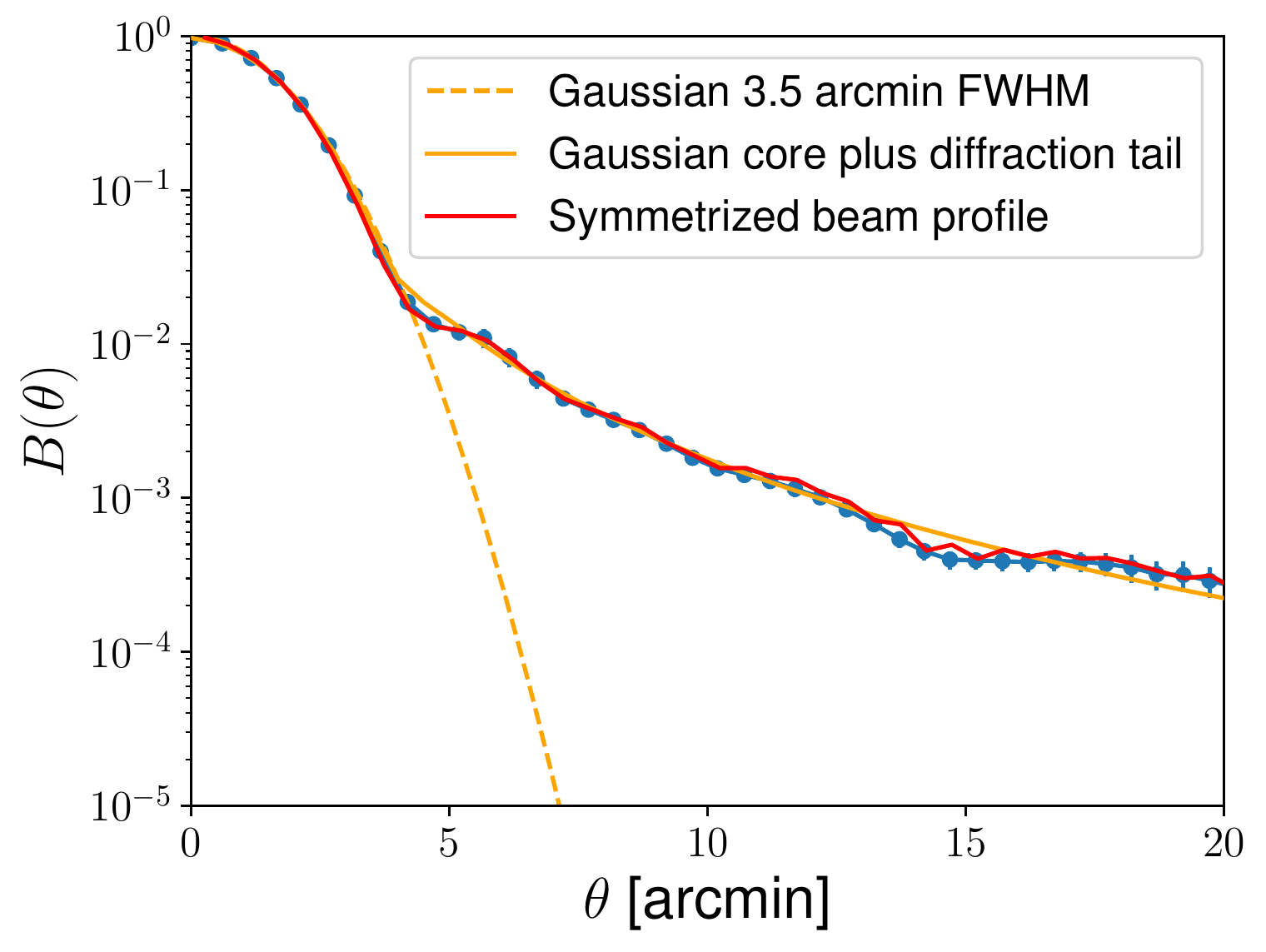}
\caption{Mean \textsc{Polarbear}\ beam profile (blue points) as a function of the the radial distance $\theta$. The mean profile has been computed as an inverse noise weighted average of all the beam profile measurements in our dataset. The error bars show the $2\sigma$ error on the weighted mean. The results for the fit of the data to a Gaussian profile and to a model including a Gaussian core and a $1/\theta^3$ diffraction tail are shown as dashed and solid line respectively. The radial profile of the beam estimated from the Jupiter maps is shown in red.}
\label{fig:beam}
\end{figure}

We estimate the beam map and its effective window function, $B_{\ell}$, using dedicated observations of Jupiter, following the procedure outlined in PB14. However, for this analysis we conservatively discarded the observations of Jupiter taken when the atmospheric precipitable water vapor (PWV) exceeded a threshold of 4~mm (as was also done for CMB observations), despite no clear sign of detector saturation in the reported range of angular scales. As a consistency check we estimated the $B_{\ell}$ for each  observation season separately to test potential variation over time and found no significant deviation. Consistent results have been derived using Saturn observations.
We tested that the deconvolution of the bolometer time constants induced negligible change on the beam properties and window function. 
The main lobes of the beam are well approximated by a Gaussian core having $3\farcm 5 \pm 0\farcm 1$ FWHM plus a diffraction tail asymptotically decaying as $1/\theta^3$ where $\theta$ is the radial coordinate of the beam profile \citep{hasselfield2013atacama}. The $1/\theta^3$ decay is a natural consequence of the presence of a Lyot stop in the optical chain as well as the finite detector size and difference in the spectral response of the detectors.
In Figure~\ref{fig:beam} we show the mean radial profile of the beam obtained as an inverse noise weighted average of the beam profiles estimated for each single Jupiter observation in our dataset together with the best fit results of the beam profile model. The difference between the $B_{\ell}$ bandpowers derived from the best fit profile model (orange curve) and the reference one derived from the Jupiter maps (red curve) is less than $1.5\%$ in all the $\ell$ range considered in this work.

The median beam ellipticity measured across the array is 5\% and the median difference in ellipticity measured for two channels in each focal plane pixel is 1.6\%, with a subdegree difference in the ellipticity orientation. In the systematics simulations of~Sec.~\ref{sec:systematic_pipeline}, we use the full distributions of those two quantities to quantify the potential biases due to beam effects on the $B$-mode power spectrum.

The Jupiter-measured beam is then symmetrized radially and convolved with a pointing error term for each of the three fields to calculate the per-field beam as a function of angular multipole that will be used in the power spectrum analysis. The profile of the symmetrized beam is consistent with the radial profile data as shown in Figure \ref{fig:beam}. 
We refer the reader to PB14 for more details on the process. 

The pointing error calculation has changed slightly from PB14, although the underlying model is unchanged. We assume the real instrumental pointing is Gaussian distributed around the reconstructed pointing with a standard deviation referred to as the pointing jitter.
In PB14, the pointing jitter and uncertainty on the jitter were calculated by fitting a beam width to subsets of the data to obtain a probability density function for the real jitter. 
However, further analysis showed that the underlying CMB anisotropies at each source's location are a significant bias in the fitted beam width, an effect not captured in the previous analysis.
{\bf Like PB14, each point source is fitted using free parameters for source position, amplitude and beam width.
The uncertainty on the parameters is then estimated from the scatter of the fitted parameters for point sources simulated using the central fitted values
against a $\Lambda$CDM{} CMB temperature background.
The pointing jitter and its uncertainty is then backed out of the fitted parameters and their uncertainties obtained from simulations.}
Using this algorithm, we find pointing jitter values of $24 \farcs 5 \pm 2 \farcs 7$, $24 \farcs 5 \pm 5 \farcs 9$, and $57 \farcs 1 \pm 6 \farcs 1$
for RA4.5, RA12, and RA23, respectively.
These jitter values are consistent with those found in the PB14 analysis, but with significantly reduced error bars due to the correct treatment to the underlying CMB fluctuations.

\subsection{Detector gains} \label{sec:gains}

We calibrate the TOD to physical Rayleigh-Jeans Kelvin temperature units  (K$_{\text{RJ}}$) in a multi-step process, following the methods of PB14. 
First we determine the relative calibration between detectors over time using a combination of an internal thermal source and Saturn observations.
With this relative calibration in hand, we can turn the TOD into a temperature map. 
Finally, we determine the absolute calibration by looking at the temperature anisotropy CMB power spectrum as discussed in Sec.~\ref{sec:absgain_and_polangle}. 
In this section, we will focus on the relative calibration process.

In brief, the relative calibration of the TOD proceeds in three steps. 
The first step combines an internal thermal calibration source (the stimulator) with the flux from Saturn. 
We discuss the complications introduced by Saturn's rings in Appendix~\ref{app:saturn_rings}. 
Next, we correct for the effect due to the polarized emission from the stimulator, which is then rotated by the HWP. 
We thus define the HWP angle-dependent template by considering observations of a variety of astrophysical sources at different HWP rotation angles.
In addition to the HWP position, the polarized response also depends on the status of the stimulator (a few hardware changes were made over the observing period) and potentially a long-term temporal drift.
As a consequence we generate a new polarization template for each of six epochs (four for season 1 and two for season 2) corresponding to changes in the stimulator hardware. 
Finally, we combine the stimulator and Saturn observations to calculate the relative gain, which is the conversion from electrical current into K$_{\text{RJ}}$\ units, for each detector and observation.

The median variation of the relative gain calibration for all detectors between two consecutive stimulator measurements is 0.5\%; we use this information for the systematic error estimate to quantify the impact  of the drift of the gains on the $B$-mode power spectrum (see Sec.~\ref{sec:systematic_pipeline}).

\subsection{Polarization angle}
\label{sec:polarization}

As in PB14, we determine the relative detector polarization angles using daily observations of Tau~A\ (the Crab nebula).
We then further improve the accuracy in the global polarization angle by nulling the $C_\ell^{EB}${} cross-spectrum (see Sec.~\ref{sec:absgain_and_polangle}), assumed to be zero.
Here we summarize this approach, with more details available in PB14. 
We also tabulate the achieved polarization angle uncertainties from this analysis, which are important to the systematics budget for the $C_\ell^{BB}${} power spectrum (see Sec.~\ref{sec:systematic_pipeline}). 

We estimate polarization angles and efficiencies by observing Tau~A{} daily for roughly 30 minutes and fitting each detector's TOD from the two seasons of data to a reference Tau~A{} map from the IRAM\footnote{Institut de Radioastronomie Millim\'etrique} 30~m telescope observation \citep{2010A&A...514A..70A} using the \textsc{Polarbear}\ Jupiter-based beam and known pointing information.
The individual-pixel angular uncertainty in each wafer is estimated to be 1$\fdg$2 using
the measured angular dispersions in each wafer. 
In addition, the total systematic uncertainties for the array-averaged and wafer-averaged angles are 0$\fdg$40 and 0$\fdg$48.
A detailed breakdown can be found in Table~\ref{tb:syserr_taua}.
{\bf By combining the two seasons of data, the systematic uncertainties from the beam and relative gain uncertainties causing temperature-to-polarization leakage are improved compared to PB14. However, the polarization angle difference between the seasons described below shows that the HWP angle uncertainty is now more significant than what was found in PB14. These changes make the total polarization angle uncertainty at the same level to that of PB14.}

The array-averaged polarization efficiency was 97.2\%, consistent with the expected value of 97.6\%  from HWP and anti-reflection coatings. 
The systematic uncertainty for the polarization efficiency is 1.6\%,
which introduces a 3.3\% multiplicative uncertainty in the CMB $B$-mode power spectrum.

\begin{figure}[!htbp]
\includegraphics[width=.5\textwidth]{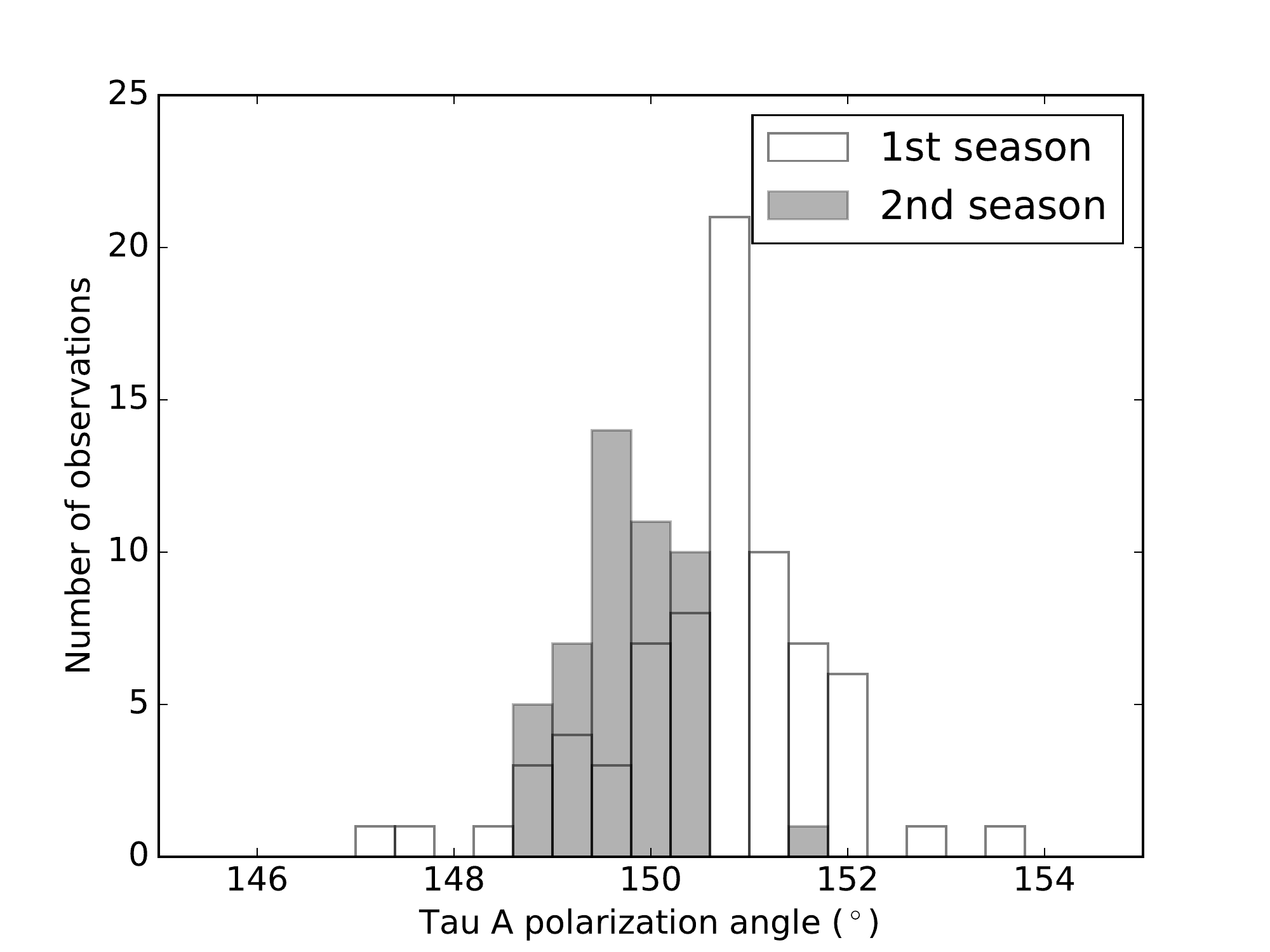}
\caption{Tau~A\ polarization angles from daily observations in the first season (unfilled) and the second season (shaded).}
\label{fig:taua_angle}
\end{figure}

We also checked the stability of the array-averaged polarization angles over time. 
Figure~\ref{fig:taua_angle} shows a histogram of the measured array-averaged Tau~A\ polarization angles. 
Each angle is calculated by computing $(1/2) \arctan \left( \sum{U_j} / \sum{Q_j} \right)$ across map pixels within 10$^\prime$ of the center of Tau~A, where $Q_j$ and $U_j$ are Stokes $Q$ and $U$ in the  map pixel $j$. 
Two effects are apparent: (1)~the average angle shifted by 0$\fdg$9 between the two seasons; (2)~the scatter was larger in season 1. 
We believe that both effects are due to the uncertainty of the HWP position. 
The polarization angle calculation is based on the commanded HWP positions (quantized at 11$\fdg$25). Any offset, $\Theta_{\rm err}$, between the commanded and actual HWP position will shift the calculated polarization angle by 2$\Theta_{\rm err}$. The first season data was taken at a number of HWP positions, (where multiple HWP offsets lead to the increased scatter in the polarization angle), while the second season data was taken at single HWP position (where the single HWP offset shows up as an polarization angle offset between first and second seasons). We can reconstruct the distribution of HWP offsets that would lead to the observed polarization angle variations, and find the typical magnitude to be 0$\fdg$28 (a polarization angle uncertainty of 0$\fdg$56).
After unblinding the $C_\ell^{EB}${} spectrum, we found that season-by-season $C_\ell^{EB}$\ nulling led to a consistent shift in the polarization angle. 
The two-season combined Tau~A\ polarization angle was $150\fdg 4 \pm 0\fdg 2 ({\rm stat}) \pm 0\fdg 8 ({\rm sys})$ in equatorial coordinates using the $C_\ell^{EB}$-derived polarization angle described in Sec.~\ref{sec:absgain_and_polangle}.

\begin{table}
\begin{center}
\caption{Uncertainties in polarization angle.}
\label{tb:syserr_taua}
\begin{tabular}{ l c c}
\hline
\hline
Angle uncertainty ($^\circ$) & Global & Wafer-Averaged \\ 
\hline
Beam uncertainties & 0.11 & 0.33\\
Relative gain uncertainties & 0.11 & 0.28 \\
Non-ideality of HWP & 0.18 & 0.19 \\
Circular polarization of Tau~A & 0.10 & 0.05 \\
HWP angle uncertainties & 0.31 & 0.04 \\
Pixel pointing uncertainties & $<$0.01 & 0.05\\
Bolometer time constant & 0.01 & 0.04\\
Filtering effect & 0.02 & 0.04 \\
\hline
Total uncertainty ($^\circ$) & 0.40 & 0.48 \\
\end{tabular}
\tablecomments{Systematic uncertainties in global reference and wafer-averaged polarization angle, as measured using Tau~A.}
\end{center}
\end{table}

\section{Data analysis}
\label{sec:data_analysis}
As mentioned, the two-season dataset contains about 61\% more data than the PB14 dataset.
This increase corresponds to new data from the second season of observation, as well as a re-analysis of the data from the first season with the improvements made in calibration and data selection, as detailed in other sections.
The main update to the data analysis with respect to PB14 has been the implementation of a second and complementary pipeline to analyze the dataset, for more robust results through consistency checks and improved systematic error control.
While the calibration of the TOD (see Sec.~\ref{sec:calibration}) is shared between the two pipelines, they perform map-making and the power-spectrum estimation differently.

\begin{figure}[htpb]
 \centering
 \includegraphics[width=3.5in]{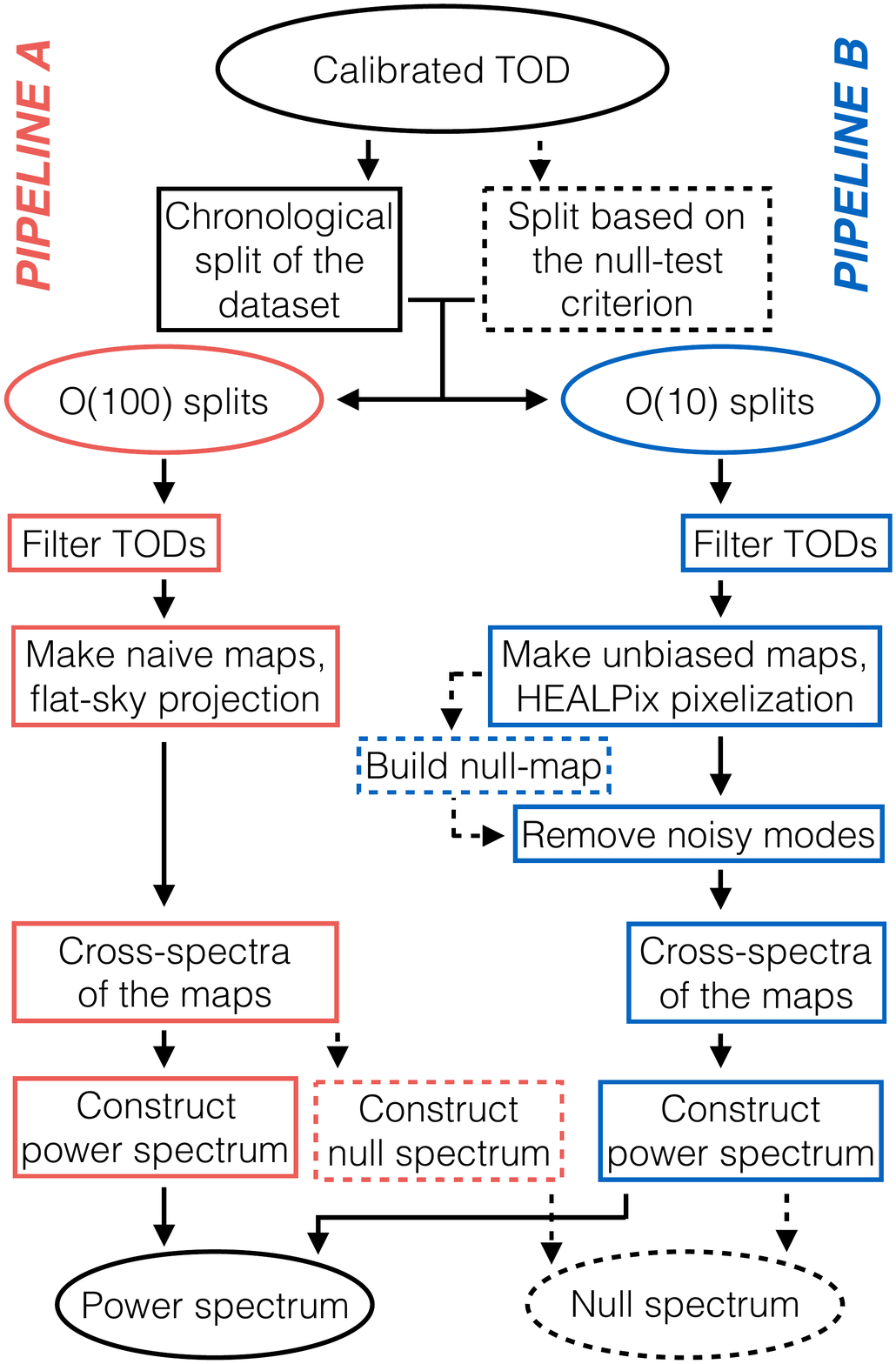}
 \caption{\label{fig:flowchart} Schematic view of the main steps performed by pipeline~A~(red) and B~(blue). The main steps starting from the calibrated TOD and leading to the production of angular cross-spectra are shown with solid boxes and arrows. In addition, extra steps related to the production of null cross-spectra used to assess the quality of the dataset are shown with dashed boxes and arrows. See text for detailed informations.
}
\end{figure}

\subsection{Map-making and power-spectra estimation}
\label{sec:data_analysis-pipelines}
We adopt two independent and algorithmically different pipelines in the analysis: the first pipeline~(hereafter ``pipeline~A'') is based on the MASTER method~\citep{Hivon2002} and was described in PB14, and the second pipeline~(hereafter ``pipeline~B'') is based on the work described in~\cite{explicit_map_making}.
The main steps performed by the two pipelines to estimate sky maps, angular cross-spectra, and null cross-spectra are described in Fig.~\ref{fig:flowchart}.

\begin{figure*}[htpb]
 \centering
 \includegraphics[width=7.5in]{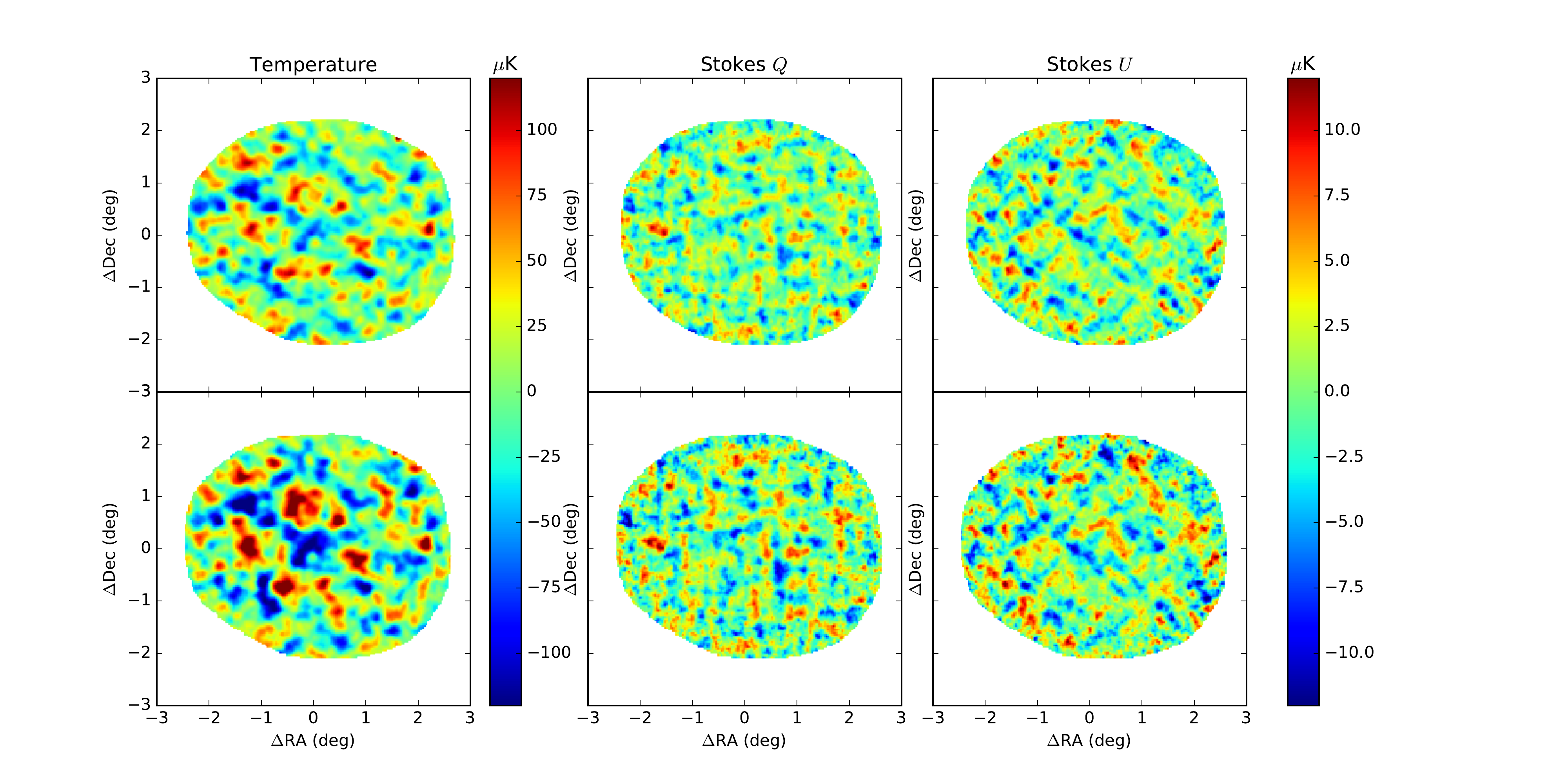}
 \caption{\label{fig:QUmaps}
\textsc{Polarbear}\ CMB intensity and polarization sky maps of RA23 in equatorial coordinates.
The left, center and right panels show temperature anisotropy, Stokes $Q$, and Stokes $U$, respectively.
The top maps are generated by pipeline~A\ and the bottom maps are generated by pipeline~B\ (resampled with a map-pixel width of $2\arcmin$ and reprojected onto a cylindrical equal-area projection to ease the comparison). Both sets of maps the maps are smoothed with a Gaussian filter with $3 \farcm 5$ FWHM, and for visualization only, we show an area in which the map weights are above -10 dB.
The polarization angle is defined with respect to the north celestial pole. 
While the structures are clearly in agreement between the two sets of maps, as expected, the amplitude of the signal is different due to the fact that the two pipelines treat the amplitudes of the modes in the maps differently~(Sec.~\ref{sec:data_analysis-pipelines} for details). Maps with alternative color schemes are available at \url{http://bolo.berkeley.edu/polarbear/data/polarbear_BB_2017/}.
}
\end{figure*}

The pipeline-A\ and pipeline-B\ map-making algorithms differ slightly in their filtering of the calibrated data and substantially in the way they project the filtered TOD into maps. 
Both perform high-pass and azimuthal filters to remove atmospheric noise and ground pickup, respectively. 
Pipeline~A\ performs these two operations sequentially while pipeline~B\ performs them simultaneously.\footnote{This is desirable in general as the templates that need to be filtered out are not always orthogonal from the outset. However,  this step makes the map-making process more complex than filtering the non-orthogonal templates one after another.}
In addition, pipeline~A\ applies a low-pass filter prior to map-making.
The pipeline-A\ map-making algorithm projects the time-domain data into maps without accounting for the filtering-induced power suppression, but accounts for this during power-spectrum estimation.
Combined with the sequential filtering, this allows a substantial speedup compared to the other pipeline.
Pipeline~B\ accounts for the filtering of the TOD and produces unbiased maps.
The two pipelines also use different pixelization schemes.
Pipeline~A\ projects the time-domain data onto flat sky maps using a cylindrical equal-area projection, with map pixels of width $2\arcmin$.
Pipeline~B\ projects the time-domain data onto curved-sky maps
using the \texttt{HEALPix}\footnote{\url{http://healpix.jpl.nasa.gov/}} pixelization~($N_{\rm side}=2048$, i.e., map pixel width $\sim 1\farcm 7$; \cite{Gorski2005}).
Given the relatively small fraction of sky observed~(a few degrees across for each patch), and the resolution of the telescope ($3\farcm 5$), both approaches remain equivalent.
In both pipelines, the data are combined into chunks which are later cross-correlated to avoid noise biases in the estimated power spectra.
After the data selection (see Sec.~\ref{sec:data-selection}), pipeline~A\ splits the two-season datasets into 214, 212, and 269 daily maps for RA4.5, RA12, and RA23, respectively;
pipeline~B\ splits the datasets into eight chunks of data for each patch corresponding to roughly 1.5 cumulative months of data each.
The full-season temperature and Stokes $Q$ and $U$ maps of RA23 produced by the two pipelines are plotted in Figure~\ref{fig:QUmaps}.
The resulting polarization white-noise levels reach 7, 6, and 5\,$\mu$K-arcmin~(10, 7, and 6\,$\mu$K-arcmin with the beam and filter transfer function divided out) for RA4.5, RA12, and RA23, respectively, for pipeline~A.

In both pipelines the power-spectrum estimators are based on the pure
pseudo-$C_\ell$ technique \citep{Smith:2006} although they differ in a few aspects, such as the computation of the mode-mode mixing matrix, and the handling of the off-diagonal elements controlling the level of $E$-to-$B$ leakage.
For pipeline~A, the formalism was described in PB14.
For pipeline~B, we use the software \texttt{X$^2$pure}\ \citep{xpure, Grain2012, ferte2013}. 
As described in \cite{explicit_map_making}, we perform a map-domain removal of the noisiest modes in order to control large-scale noise.
Figure~\ref{fig:QUmaps} shows the effective maps after performing this mode removal.
We estimate the statistical uncertainty on the $C_\ell^{EB}${} and $C_\ell^{BB}${} spectra from 500 signal and noise Monte Carlo\ simulations (100 for pipeline~B) as described in Sec.~\ref{sec:result-cmb}. 
For pipeline~A, thanks to a realistic uncertainty estimate as well as overall larger dataset and improved calibration, this release achieves almost twice the sensitivity of PB14.
As a result of applying different treatments to the data,
the estimated statistical uncertainties from the two pipelines are different~(see \cite{explicit_map_making} for a complete comparison of methods),
being on average 20\% higher for pipeline~B.
Part of this difference is also attributed to the difference of resulting sky area used in the two analyses.

In PB14, \textsc{Polarbear}{} implemented a blind analysis method in our measurement of the $B$-mode polarization, and we have adopted the same procedure for the two-season analysis.
In the following, we describe the data selection, analysis validation by a null-test framework, instrumental systematic-error estimation, and foreground estimation including biases from contamination due to point-like extra-Galactic and polarized Galactic foregrounds.

 \subsection{Data selection}
\label{sec:data-selection}

In this section, we describe the data quality cuts. 
As mentioned in Sec.~\ref{sec:data}, 60\% of the total CMB observation data (2{,}800 out of 4{,}700 hours) ends up being included in the science analysis of this work. 
We do not select the other 40\% due to some combination of bad weather, incomplete observations, or hardware glitches. 
Note that the thresholds for the latter were validated and finalized while running the suite of null tests described in the next section, and before unblinding.

The data quality checks proceed in three stages. 
First, we require a successful measurement of the detector gains for each CES (see Sec.~\ref{sec:gains} for details on how detector gains are estimated). 
Second, we discard data based on conditions such as the level of PWV in the atmosphere, the angle between the observed patch and the Sun or the Moon, as well as measures of the quality of the data, such as the bolometer yield being too low, problematic scan length, bad azimuth encoder data, or the array median gain changing too rapidly, for example.
This step can remove data from an entire CES or some of the detector channels within a CES, {\bf leaving $\sim$ 10,000 and $\sim$4,000~CES} for season 1 and season 2 respectively. 
Note that, thanks to the improved calibration and slightly different data selection, the definition of season 1 here is not strictly equivalent to that used in PB14; the change increased by approximately {\bf 15\% of the observation-hours} that went into the maps.
In the end, 2{,}800~observation-hours pass both the first and second data quality checks.
{\bf The volume of data from the second season is smaller than the one from first season mainly because the period of observing time is smaller (13 months for first season and 6 months for second season).}
The third step goes further and consists of defining subscans (a constant-velocity segment of a timestream) and among them, identifying contaminated subscans (such as by finding glitches in them).
{\bf As in PB14, we discard all data obtained while the telescope is accelerating.}

 \subsection{Analysis validation: null tests}
\label{sec:sys-null}

We perform a suite of null tests to evaluate the calibration, data-selection criteria and filtering methods, and to test for unknown systematic errors, before unblinding the data. 
This task consists of iteratively running the null-test framework described in PB14, with the addition of three new data splits, until a set of predefined criteria (described below) are passed.
Due to the non-negligible extra computational cost of pipeline~B\ with respect to pipeline~A,\footnote{A single map-making run of pipeline~B\ uses roughly as much computation time as one full run of the null-test framework of pipeline~A.} the full set of null tests (see Sec.~\ref{sec:null-data-splits}) is performed only by pipeline~A\ as detailed in PB14. We primarily used these null tests by pipeline~A\ to define the data selection criteria. Once the science dataset is defined, pipeline~B\ performs a subset of the data splits probed by pipeline~A.
The way pipeline~B\ computes null-spectra differs slightly from pipeline~A. It first splits the dataset in two according to the null-test criterion; second, it computes unbiased maps of 4 (or 8) disjoint subsets of each split; null-maps are then computed by taking the difference of maps that belong to different splits; and finally it computes and co-adds the cross-spectra of the null maps to form a null spectrum. Nevertheless, pipeline~B\ performs further checks allowed by the production of null maps in addition to null spectra. These include the visual inspection of null maps, the study of the individual cross spectra of null maps, and their distribution (the same data can contribute to several cross-spectra). 
More specifically, if the band powers of a null spectrum have a probability-to-exceed~(PTE, see also Sec.~\ref{sec:null-test-analysis}) less than 5\%, we study the cross-spectra of which it is the average and we make sure that the low PTE is not determined by the cross-spectra involving the same particular portion of the data set.
None of those extra checks found signs of contamination or inconsistencies in the dataset.
We emphasize that this large variety of the validation tests is made possible by combining the complementary strengths of the two pipelines.

\subsubsection{Data splits}\label{sec:null-data-splits}

The null tests are performed for several interesting splits of the data, chosen to be sensitive to various sources of systematic contamination or miscalibration.
In addition to the null tests we performed in PB14~(``first half versus second half of the dataset'', ``rising versus setting'', ``high elevation versus low elevation'',
``High gain versus low gain'', ``Good versus bad weather'',
``pixel type'',\footnote{Each detector wafer has two different pixel polarization angles.}
``left versus right side of the focal plane'',
``left- versus right-going subscan'' and ``Moon distance''), we introduced the following three new tests to check the difference between seasons
and our possible concerns:
\begin{itemize}
\item ``First season versus second season'': probing seasonal variation on year-long time-scales.
This test is sensitive to systematic changes in the calibration, beams, telescope, and detectors.

\item ``Sun distance'': checking for residual contamination after setting the Sun-proximity threshold for an observation to be considered for analysis.

\item ``Sun above the horizon versus sun below the horizon'': checking for contamination from the far sidelobe of the beam,
and systematic changes of the pointing due to the small deformation of the telescope by solar heating.
\end{itemize}
The 12 null tests are used to analyze the dataset, and the correlations between tests are taken into account in the analysis by also running the same suite of null tests on Monte Carlo (MC) simulations as described below. For the 3 sky patches, pipeline~A\ runs then a total of 36 null tests, while pipeline~B\ focuses on a subset of 11 null tests.

\subsubsection{Analysis} \label{sec:null-test-analysis}
\begin{figure}[htbp]
 \centering
 \includegraphics[width=3.4in]{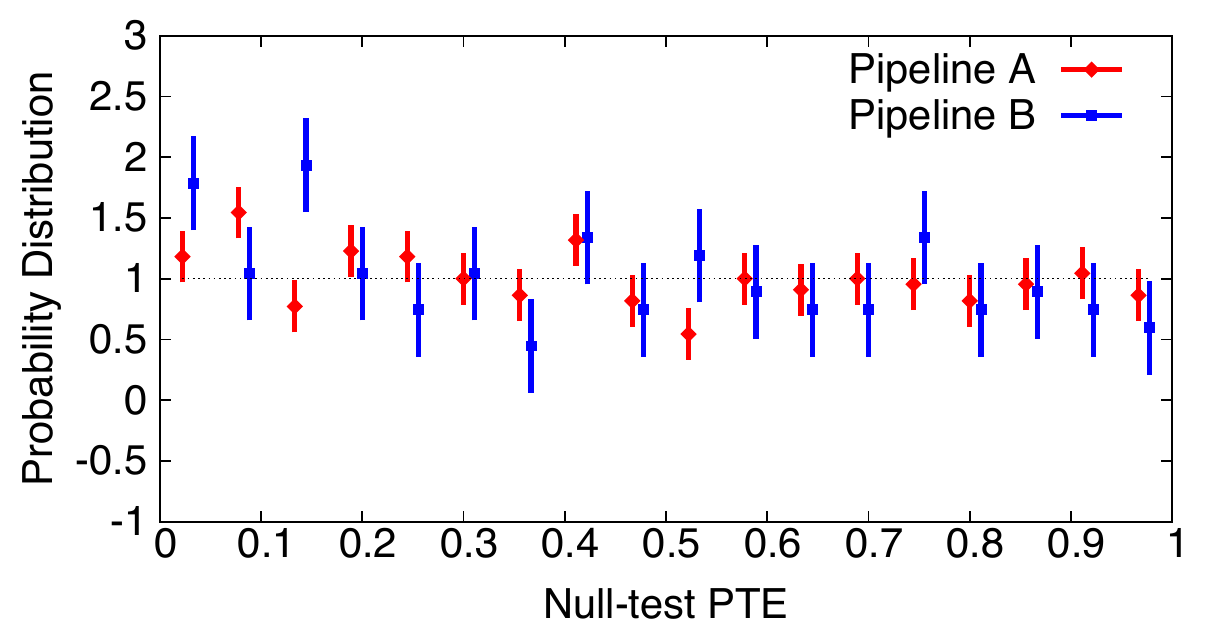}
 \caption{Null-test-PTE distribution for $\chi^2_{\rm null}$ for both pipeline~A~(red) and pipeline~B~(blue) (396 and 121 entries respectively). Both distributions are consistent with the expected uniform distribution.}
 \label{fig:dist_chi2}
\end{figure}

In the null spectra, for each band power bin~$b$, we calculate the statistic $\chi_{\rm null}(b)\equiv \hat{C}_b^{\rm null}/\sigma_b$,
where $\sigma_b$ is an MC-based estimate of the corresponding standard deviation, and its square $\chi_{\rm null}^2(b)$.
$\chi_{\rm null}(b)$ is sensitive to systematic biases in the null spectra, while $\chi_{\rm null}^2(b)$ is more sensitive to outlier bins. 
To probe for systematic contamination affecting a particular power spectrum or null-test data split,
we calculate the sum of $\chi_{\rm null}^2(b)$ over $500 < b < 2100$ by spectrum~(``$\chi^2_{\rm null}$ by spectrum''),
and the sum for both spectra for a specific test~(``$\chi^2_{\rm null}$ by test'').
We consider both $C_\ell^{EB}${} and $C_\ell^{BB}${} in the null tests in order to investigate sources of spurious $B$-mode signals.
We require that each of these sets of PTEs are consistent with a uniform distribution,
as evaluated using a Kolmogorov-Smirnov test,
requiring a $p$-value~(probability of seeing deviation from uniformity greater than that which is observed given the hypothesis of uniformity) to be equal to or greater than 5\%.
These distributions are consistent with the uniform distribution
and Figure~\ref{fig:dist_chi2} shows the PTE distribution of the $\chi^2_{\rm null}$ for the two pipelines.

We create test statistics based on these quantities to search for different manifestations of systematic contamination. 
The five test statistics are PTEs from (1)~the average value of $\chi_{\rm null}$; the extreme value of $\chi^2_{\rm null}$
(2)~by bin, (3)~by spectrum, and (4)~by test; 
(5)~the total $\chi^2_{\rm null}$ summed over the twelve null tests. 
In each case, the result from the data is compared to the result from simulations, and PTEs are calculated.
Finally, we combine each of the test statistics, and calculate the final PTE, requiring it to be equal to or greater than 5\%. 
Table~\ref{tab:pte_summary} shows the PTE values for the final configuration.

Comparing the most significant outlier from the five test statistics with that from signal and noise simulations,
pipeline~A~(pipeline~B) gets PTEs of 71.8\%~(77\%), 65.2\%~(16\%), and 16.6\%~(13\%) for RA4.5, RA12, and RA23 respectively.
We therefore achieve the requirements described above, finding no evidence for systematics or miscalibration in the \textsc{Polarbear}{} dataset used for the analysis and in the analysis process itself. 
\begin{table*}[htbp]
\begin{center}
\caption{\label{tab:pte_summary}PTEs resulting from the null test framework.}
\begin{tabular}{c c cc c cc c cc c cc c cc}
\tableline
\tableline
 Patch && \multicolumn{2}{c}{\shortstack{Average of\\ $\chi_{\rm null}(b)$}}
       && \multicolumn{2}{c}{\shortstack{Extreme of\\ $\chi_{\rm null}(b)$}}
       && \multicolumn{2}{c}{\shortstack{Extreme of\\ $\chi^2_{\rm null}$ by $C_\ell^{EB/BB}$}}
       && \multicolumn{2}{c}{\shortstack{Extreme of\\ $\chi^2_{\rm null}$ by test}}
       && \multicolumn{2}{c}{\shortstack{Total \\ $\chi^2_{\rm  null}$}} \\
\cline{3-4}
\cline{6-7}
\cline{9-10}
\cline{12-13}
\cline{15-16}
       && \multicolumn{2}{c}{Pipeline}
       && \multicolumn{2}{c}{Pipeline}
       && \multicolumn{2}{c}{Pipeline}
       && \multicolumn{2}{c}{Pipeline}
       && \multicolumn{2}{c}{Pipeline} \\
       && A & B && A & B && A & B && A & B && A & B \\
\tableline
 RA4.5 && 84.0\% & 100\%           && 70.8\% & 49\% && 86.0\% & 54\% && 64.0\% & 73\% &&           35.4\% &           47\% \\
 RA12  && 70.8\% & \phantom{10}9\% && 29.8\% & 19\% && 66.2\% & 52\% && 57.8\% & 54\% &&           51.0\% &           52\% \\
 RA23  && 95.0\% & \phantom{1}91\% && 40.8\% & 13\% && 69.4\% & 11\% && 42.8\% & 16\% && \phantom{0}5.0\% & \phantom{0}5\% \\
\tableline
\end{tabular}
\tablecomments{PTEs resulting from the null-test framework.
No significantly low or high PTE values are found, consistent with a lack of systematic contamination or miscalibration in the \textsc{Polarbear}\ dataset and analysis. 
The only exception is the average $\chi_{\rm null}(b)$ of RA4.5 for pipeline~B.
The 100\% PTE results from all the 100 noise-only simulations having worse
$\chi_{\rm null}(b)$ than the data. The possibility that this is caused by 
 overestimation of the error bars is, however, excluded: this would produce a distribution of the
PTEs skewed high (in contrast with~Figure~\ref{fig:dist_chi2}).
Note that the PTE values in each patch are not independent from one another.}
\end{center}
\end{table*}
 \subsection{Analysis validation: simulations of instrumental effects} \label{sec:systematic_pipeline}

We describe in this section the signal-only simulations used to determine the effect of  uncertainties in the instrument model on the power spectrum listed in Table~\ref{tab:summary_systematic_uncertainties} (sub-section ``Instrument'').
We investigate nine systematic instrumental effects: uncertainty in instrument polarization angle;  uncertainty in relative pixel polarization angles; uncertainty in instrument boresight pointing model; differential pointing between the two detectors in a pixel; the drift of the gains between two consecutive thermal source calibrator measurements; relative gain calibration uncertainty between the two detectors in a pixel; crosstalk in the multiplexed readout; differential beam size; and differential beam ellipticity. 

The pipeline used to analyze those systematic effects was described in PB14, while here we describe two improvements: (1) the systematics pipeline follows exactly all the data analysis pipeline steps, and (2) all instrumental systematics studied are now included within the systematics pipeline.
The first modification makes the systematic error study more comprehensive than in PB14,
i.e., the simulations performed here also include time-domain filtering and the cross-correlation of submaps.
The second modification allows us to have a common framework for all effects, and therefore the same metric to analyze the results (i.e., we now include beam effects and electrical crosstalk, which were analyzed separately in PB14).
While the procedure was developed for both data analysis pipelines, we report here only the results from pipeline~A\ due to computing constraints.

A major change in the hardware configuration with respect to PB14 is the position of the HWP. 
As mentioned in Sec.~\ref{sec:polarization}, we recorded the data beginning from the middle of the first season without stepping the HWP, in order to decrease the scatter in the polarization angle of the detector.
We found that this results in less mitigation of some instrumental effects such as those related to instrument and pixel polarization angle, drift of the gains, or crosstalk in the multiplexed readout.
We note, however, that pointing-related effects remain stable with respect to PB14, and we reduce the uncertainty coming from beam-related effects mainly thanks to an improved characterization.

\begin{figure}[htbp]
 \centering
 \includegraphics[width=3.4in]{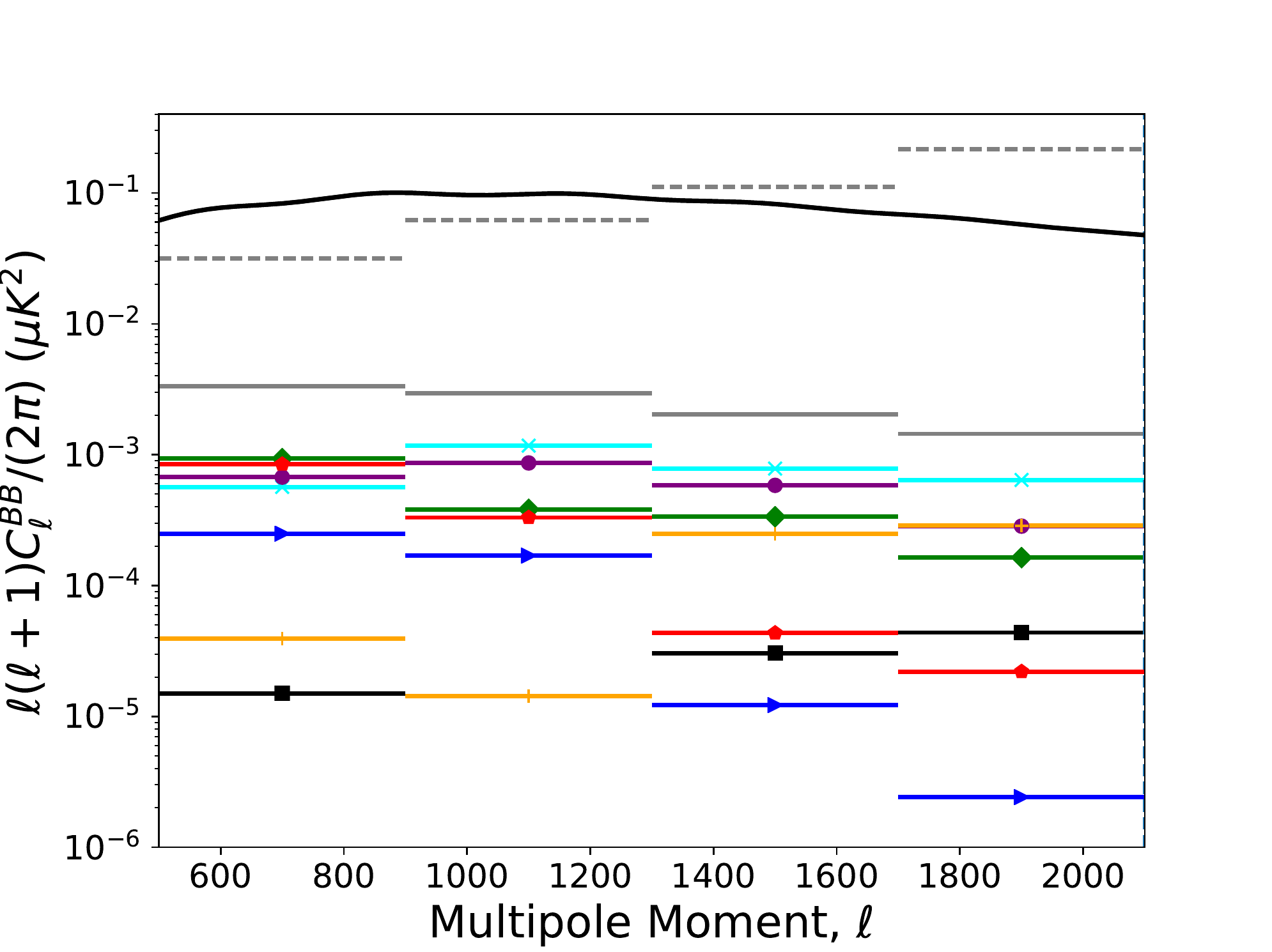}
 \caption{\label{fig:all_systematics}
Estimated levels or upper bounds on instrumental systematic uncertainties in the four bins of the $C_\ell^{BB}${} power spectra, as described in Sec.~\ref{sec:systematic_pipeline}. Individual effects (solid colors) and their combination (solid horizontal grey line) are displayed:  combined uncertainty in instrument polarization angle and relative pixel polarization angles after self-calibration (purple circle), combined uncertainty in instrument boresight pointing model and differential pointing between the two detectors in a pixel (cyan cross), the drift of the gains between two consecutive thermal source calibrator measurements (red star), relative gain-calibration uncertainty between the two detectors in a pixel (green diamond), crosstalk in the multiplexed readout (blue arrow), differential beam shape (orange plus), and differential beam ellipticity (black square). For comparison we display the binned statistical uncertainty from the pipeline~A{} (dashed horizontal line) reported in Table~\ref{tab:bandpowers} and the theoretical Planck~2015\ $\Lambda$CDM\ lensing $B$-mode spectrum (solid black line).}
\end{figure}

All nine effects and their combination were found to produce spurious BB power well below the statistical uncertainty in the measurement of $C_\ell^{BB}$, as also shown in Fig.~\ref{fig:all_systematics}. 
The individual effects are combined linearly to give the total contamination.

\section{Foregrounds}
\label{sec:sys-fg}
\begin{table*}
 \begin{center}
 \caption{\label{tab:fgs} Sources of foreground power and their expected power in $D_\ell^{BB}$.}
 \begin{tabular}{lcccc}
 \hline
 \hline
 Foreground & \multicolumn{4}{c}{Expected power in $D_\ell^{BB}$~($10^{-4}$ $\mu$K$^2$)}\\
 \cline{2-5}
 & $\ell=500\text{--}900$ & 900--1300 & 1300--1700 & 1700--2100 \\ 
 \hline
 Galactic dust           & $63.5 \pm 123.3$                     & $53.6 \pm 102.4$                     & $48.8 \pm 91.4$                     & $\phantom{0}46.9 \pm 85.1$ \\
 Galactic synchrotron    & $\phantom{0}1.4 \pm \phantom{00}2.1$ & $\phantom{0}1.2 \pm \phantom{00}1.9$ & $\phantom{0}1.1 \pm \phantom{0}1.7$ & $\phantom{00}1.0 \pm \phantom{0}1.5$ \\
 Radio \& dusty galaxies & $13.4 \pm \phantom{00}5.5$           & $26.8 \pm \phantom{00}9.8$           & $24.8 \pm 15.3$                     & $\phantom{0}67.4 \pm 21.9$ \\
 \hline
 Total & $78.3 \pm 123.4$ & $81.6 \pm 102.9$ & $74.7 \pm 92.7$ & $115.3 \pm 87.9$ \\
 \hline
 \end{tabular}
  \tablecomments{
  The total central value~(uncertainty) on the final line is the linear~(quadrature) sum of the individual foreground powers. Note that reported values for Galactic dust and synchrotron are upper limits and we have no detection of dust contamination nor of synchrotron contamination in our observed fields.}
 \end{center}
\end{table*}

Polarized Galactic and extra-Galactic foregrounds are a potential contaminant to the CMB, and a particular concern for the very faint $B$-modes. 
As described in PB14, there are four foregrounds sources of interest: 
the first two are polarized Galactic dust and synchrotron emission, dominating at large angular scales (down to a few arcminutes) at intermediate and high Galactic latitudes; the second two are emission from polarized radio and dusty galaxies, on scales of a few arcminutes scale and smaller.
In this section, we describe how we estimate the band power contribution from each of these sources.
Note that the methods employed to estimate the polarized Galactic dust and synchrotron contamination is different from the one described in PB14, while the method used to estimate contamination from radio and dusty galaxies remains similar.
Table~\ref{tab:fgs} reports the mean values and 68.3\% confidence intervals obtained for the band powers from the combination of the three patches for both components, and Figure~\ref{fig:all_foregrounds} shows the estimated foreground contributions to $\ell \left ( \ell + 1 \right ) C_\ell^{BB} / 2\pi$\ at a 68.3\% confidence level.
As can be seen, the total foreground contribution is estimated to be small in any band power, though not completely negligible. 

{\bf Note that the methods employed to estimate the polarized Galactic dust and synchrotron contamination is different from the one described in PB14,
while the method used to estimate contamination from radio and dusty galaxies remains similar.
Indeed, in PB14 the estimates of diffuse polarized foregrounds in the observed fields was based on models built upon the public data
from {\sc wmap}{} and Planck at the time for the dust~\citep{2013ApJS..208...20B, 2014A&A...571A...1P},
as well as the QUIET levels reported for synchrotron~\citep{2012ApJ...760..145Q}. 
Between the PB14 and this paper, the Planck~2015{} polarization sky maps, including in particular the 353~GHz as the main tracer of the polarized Galactic dust emission,
have become available~\citep{2016A&A...594A...1P}.\footnote{\url{https://www.cosmos.esa.int/web/planck/pla}} 
We therefore use this new data as the main reference for both the dust and synchrotron estimates reported below.
It is relevant to notice also that, while carrying out the present analysis, we discovered a conversion error in the evaluation of the contamination from synchrotron emission in PB14,
which however does not change significantly the conclusions there, and that has been reported in a separate {\it erratum}.}

\subsection{Polarized Galactic dust \& synchrotron}
\label{sec:sys-fg-diffuse}
The polarized Galactic foregrounds, synchrotron and thermal dust, dominate at frequencies smaller and larger than 70~GHz, respectively~\citep{2016A&A...594A..10P};
at intermediate and high Galactic latitudes, and at the degree angular scale, the frequency of foreground minimum seems to vary substantially,
almost equally distributed between 60 and 90~GHz in the regions where both foregrounds are detected with high significance~\citep{2016A&A...588A..65K}.
These Galactic foregrounds substantially contaminate CMB $B$-mode measurements at all frequencies on large angular scales,
even in the cleanest regions of the sky, as shown in several recent studies~\citep{2016A&A...586A.133P, 2016A&A...588A..65K}.  
Nevertheless, Galactic foregrounds are expected to be sub-dominant with respect to the lensing $B$-modes at arcminute scales, and our purpose here is to assess their relevance and provide upper limits on their contribution to our observations. \par

For thermal dust, we adopt the following procedure. We take the publicly available Planck~2015\ sky map at 353 GHz as a tracer of polarized emission from thermal dust. To avoid noise bias in the computation of spectra, we calculate cross-spectra using half-mission (HM) splittings using the \texttt{X$^2$pure}\ power spectrum estimator. We evaluate statistical errors by means of white noise MC simulations, using the pixel-pixel noise covariance matrices of the input maps. The low signal-to-noise ratio of Planck~2015\ polarization maps at high Galactic latitudes prevents the estimation of $B$-mode spectra directly in the regions corresponding to our patches, because of their reduced size and the noise level at the small angular scales we consider. Therefore, in order to provide an upper limit on the amplitude of thermal dust, we compute power spectra on larger circular regions, with $10^{\circ}$ radius,
centered on our sky patches. 
We extrapolate the measured amplitude of the spectra at $\ell\simeq80$ (a multipole bin between 60 and 99) in these regions to higher multipoles by applying the power law scaling ${\cal D}_{l}\propto\ell^{\alpha_d}$ with $\alpha_d=-0.42\pm0.02$ \citep{2016A&A...586A.133P}. 
We then scale the Planck~2015\ measurements to the \textsc{Polarbear}\ frequency assuming a modified blackbody spectral dependence for the thermal dust, with temperature $T_{d}\simeq 19.6$ K and $\beta_{d}\simeq 1.59\pm0.14$ \citep{planck2014-XXII}. 
For computing the extrapolation in frequency we take into account both the Planck~2015\ and the \textsc{Polarbear}{} frequency band passes, and for the extrapolation in $\ell$ we consider the actual \textsc{Polarbear}\ band power window function. To account for the fact that we are measuring the foreground amplitude on larger regions, we include in the error budget the Gaussian approximation of the signal sample variance evaluated for the actual sky area of the \textsc{Polarbear}{} patches. \par
A similar procedure has been used to estimate the amplitude of polarized synchrotron  emission. We computed power spectra for synchrotron $B$-modes by cross-correlating the Planck~2015-Low Frequency Instrument 30 GHz map at the effective frequency of 28.4 GHz \citep{2016A&A...594A...1P} with the {\sc wmap}-K map at a frequency of 22.8\,GHz \citep{2013ApJS..208...20B}. Spectra are computed on the same circular regions of $10^{\circ}$ radius. The measured amplitude at $\ell\simeq 80$ is rescaled in frequency considering a power-law frequency dependence with $\beta_s=-3.12\pm0.02$ \citep{Fuskeland14} and in multipole considering ${\cal D}_{l}\propto\ell^{\alpha_s}$ with $\alpha_s=-0.31\pm0.13$ \citep{2016A&A...594A..10P}.  

\begin{figure}[htbp]
 \centering
 \includegraphics[width=3.4in]{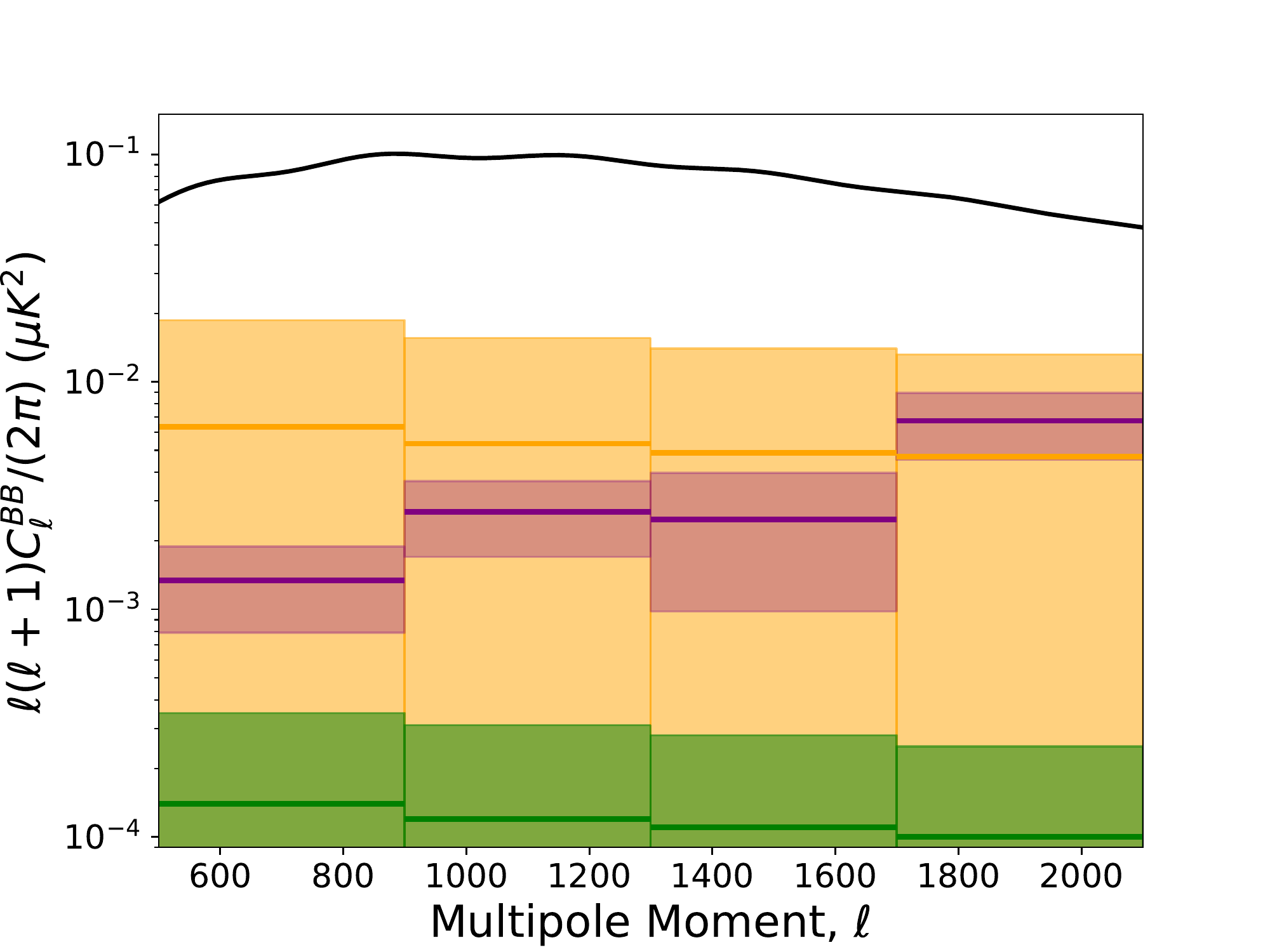}
 \caption{\label{fig:all_foregrounds} Estimated foreground contributions to $\ell \left ( \ell + 1 \right ) C_\ell^{BB} / 2\pi$\ at 68.3\% confidence intervals in the multipole range $500 \leq \ell \leq 2100$: the upper limits on polarized Galactic foregrounds, synchrotron (green shaded area) and thermal dust (orange shaded area), and the combined radio and dusty power (purple box). {\bf The solid horizontal lines represent the mean value in each band power (see Table~\ref{tab:fgs}).} As can be seen, the foreground contributions are small, although not completely negligible. For comparison, a theoretical Planck~2015{} $\Lambda$CDM\ spectrum (solid black line) is shown.}
\end{figure}

We note that for both polarized Galactic dust and synchrotron, the limiting factor in our calculation of the foreground mean values and uncertainties comes from the Planck~2015\ polarization noise level at such small angular scales. {\bf The level of contamination reported in this paper is therefore higher than our previous PB14 estimates,
which was based on available models at the time. In particular, the dominant contribution from Galactic dust,
represented by the top of the orange area in Figure~\ref{fig:all_foregrounds}, raises by a factor of about 5 with respect to PB14.}

\subsection{Radio \& dusty galaxies}
\label{sec:sys-fg-radio_and_dusty_galaxies}

We estimate the radio and dusty galaxy power by drawing 10,000 realizations from the distributions described below. 
In both cases, we take recent measurements of the temperature power and then convert to polarization power using an estimate of the mean square polarization fraction. 
For radio galaxies, the temperature power estimate comes from the \citet{dezotti} model at 150\,GHz, which is scaled to the \textsc{Polarbear}{} effective frequency according to the measured spectral index for the radio sources of $\alpha_{\rm rg} =-0.90\pm0.20$  from \citet{2015ApJ...799..177G} (hereafter G14). 
We ignore shot noise due to finite sky area, but assume a 10\% calibration uncertainty on our nominal masking threshold of 25\,mJy (i.e., the modeled threshold is $25\pm2.5$\,mJy). 
We take the recent estimate from {\sc SPTpol}\ of the mean square polarization fraction for synchrotron sources of $\left< p^2 \right> = (1.42\pm0.15)\times 10^{-3}$ (T.~Crawford, {\it private communication}).\footnote{\url{https://cmb-s4.org/CMB-S4workshops/index.php/File:Sptpol_ptsrc_polfrac_500d.pdf}} 
These numbers are consistent with a recent analysis of brighter synchrotron sources in the 143\,GHz Planck~2015{} maps which found $\sqrt{\left< p^2 \right>} = 0.043\pm 0.0018$, corresponding to $\left< p^2 \right> = (1.85\pm0.15)\times 10^{-3}$ \citep{bonavera17}. 

We use direct observations of the dusty galaxy power in temperature from G14.
These include a Poisson term with $D_{3000}^{P} = 9.16\pm0.36$\,$\mu$K$^2${} at the {\sc spt}\ frequency (150\,GHz) and a spectral index of $\alpha_P=3.267\pm0.077$, and a clustered term of the form $\ell^{0.8}$ normalized to $D_{3000}^{C} = 3.46\pm0.54$\,$\mu$K$^2${} with a spectral index of  $\alpha_C=4.27\pm0.2$. 
Data on the polarization fraction of these dusty galaxies are still poor, but they are expected to have lower polarization fractions than synchrotron sources. 
We conservatively draw the mean square polarization fraction $\left< p^2 \right>$ from a uniform distribution between $1\times10^{-4}$ and $1.57\times10^{-3}$ (i.e., 1\% polarized to the +1 sigma limit from the G14 synchrotron measurement). 

For each realization, we multiply the inferred dusty and radio galaxy spectra by the appropriate window functions to directly compare to the measured band powers.

\section{Calibration using CMB spectra} \label{sec:absgain_and_polangle}

The absolute gain calibration is performed differently by the two pipelines.
Pipeline~A\ combines the estimate of $C_\ell^{TT}$\ from each patch into one single estimate according to their statistical and beam uncertainties, and then estimates the absolute gain by fitting the patch-combined $C_\ell^{TT}$\ to a theoretical Planck~2015{} $\Lambda$CDM\ spectrum.\footnote{All references to the Planck~2015{} $\Lambda$CDM\ model in this work refer to the best-fit values for the {\tt{base$\_$plikHM$\_$TT$\_$lowTEB$\_$lensing}} configuration.} Finally the correction is applied to the individual maps.
Pipeline~B\ estimates the absolute gain for each patch by first cross-correlating the \textsc{Polarbear}\ temperature maps with the foreground-cleaned Planck~2015\ temperature map produced by the Spectral Matching Independent Component Analysis (SMICA) method ~\citep{2016A&A...594A...9P}. 
These cross-spectra are then compared to the corresponding \textsc{Polarbear}\ auto-spectra, before combining the estimates into a single absolute gain factor.

Both methods give consistent results, however the second method gives a higher uncertainty due to the propagation of the noise in Planck~2015\ data: 3.0\% uncertainty on the absolute gain factor for pipeline~A, and 4.3\% uncertainty for pipeline~B.
We note, however, that the method used by pipeline~B\ has the advantage of being more cosmology independent.
These calibration factors are then applied to all spectra by the respective pipelines to produce gain-calibrated spectra. 
The beam uncertainty is estimated from uncertainty in the point-source-derived beam-smoothing correction~(Sec.~\ref{sec:beam}),
and the variation in that correction across each field.
We shift the simulated beam by $1\sigma$, and find this beam shift leads to a $\pm 1.0\%$ change in the best-fit lensing $B$-mode amplitude.
We include this $1.0\%$ uncertainty as part of the multiplicative error budget.
A complete breakdown of the multiplicative uncertainties can be found in Table~\ref{tab:summary_systematic_uncertainties}.

The global instrument polarization angle correction $\Delta \psi$ is obtained by fitting the patch-combined $C_\ell^{EB}${} to $2 \Delta \psi C_\ell^{EE}$~\citep{ksy2013} for both pipelines.
The best-fit value and statistical uncertainty in the global $C_\ell^{EB}$-derived instrument polarization angle correction is $-0\fdg79\pm 0\fdg16$~($-0\fdg67\pm 0\fdg17$) for pipeline~A~(pipeline~B).
Combining this value with the relative shift from the Tau~A-derived angle (see Sec.~\ref{sec:polarization}), this result is consistent with the results obtained previously in PB14.
Finally, pipeline~A{} applies the polarization angle correction to the individual maps and pipeline~B{} to the power spectra.

\section{$B$-mode power spectrum results}
\label{sec:result-cmb}

\begin{figure} \centering
 \includegraphics[width=3.45in]{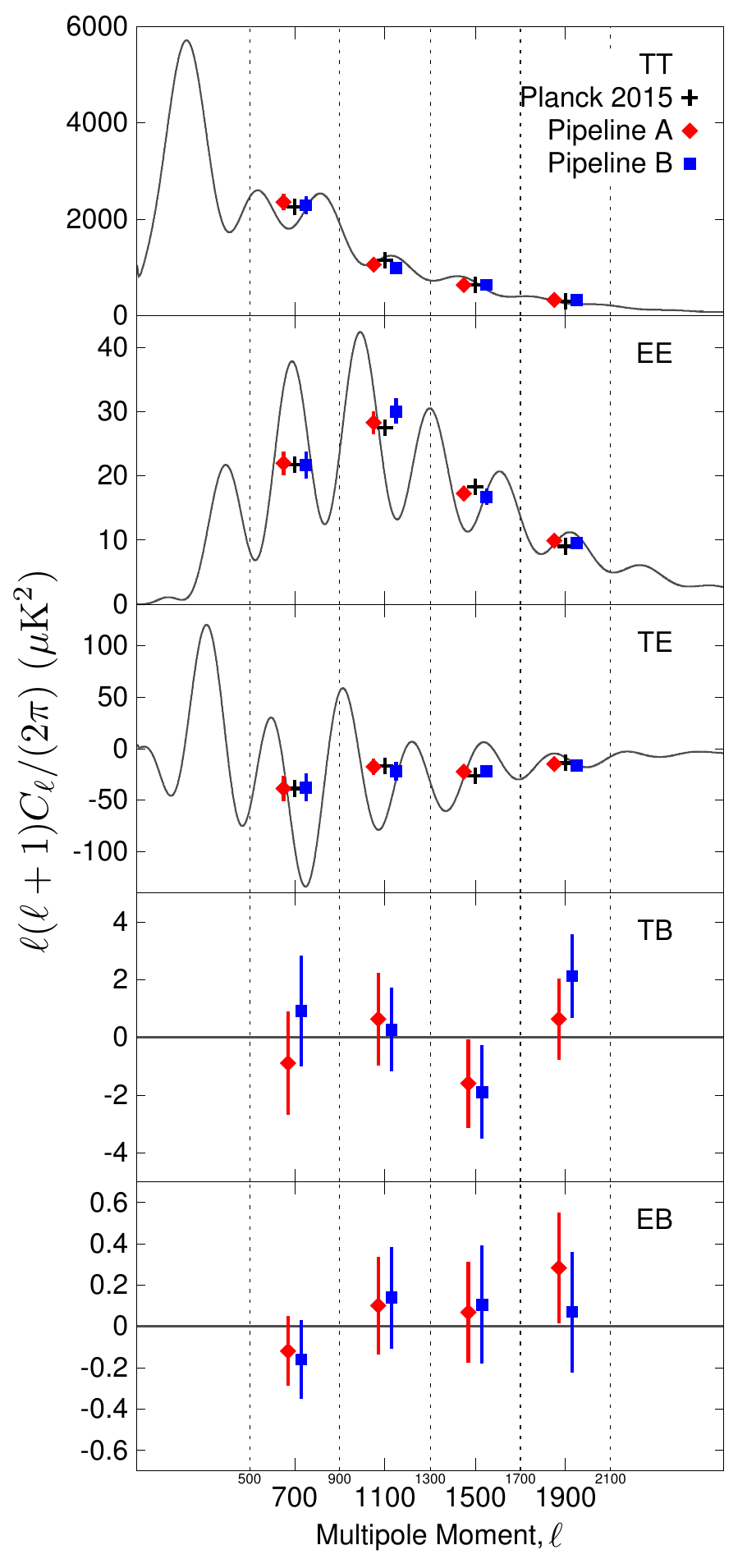}
 \caption{\label{fig:spectrumresults}\textsc{Polarbear}\ power spectra from the two-season datasets used for calibration and cross-checks. 
Red diamonds~(blue squares) show the measured band powers from pipeline~A~(pipeline~B). The uncertainty shown for the band powers is the diagonal of the band power covariance matrix, including beam covariances, and all results take into account the absolute gain factor and global instrument polarization angle (see Sec.~\ref{sec:absgain_and_polangle}).
The black curve is a theoretical Planck~2015{} $\Lambda$CDM\ spectrum, and the black pluses are the expected binned band powers.}
\end{figure}
\begin{figure*}[htpb]
 \centering
 \includegraphics[width=5.5in]{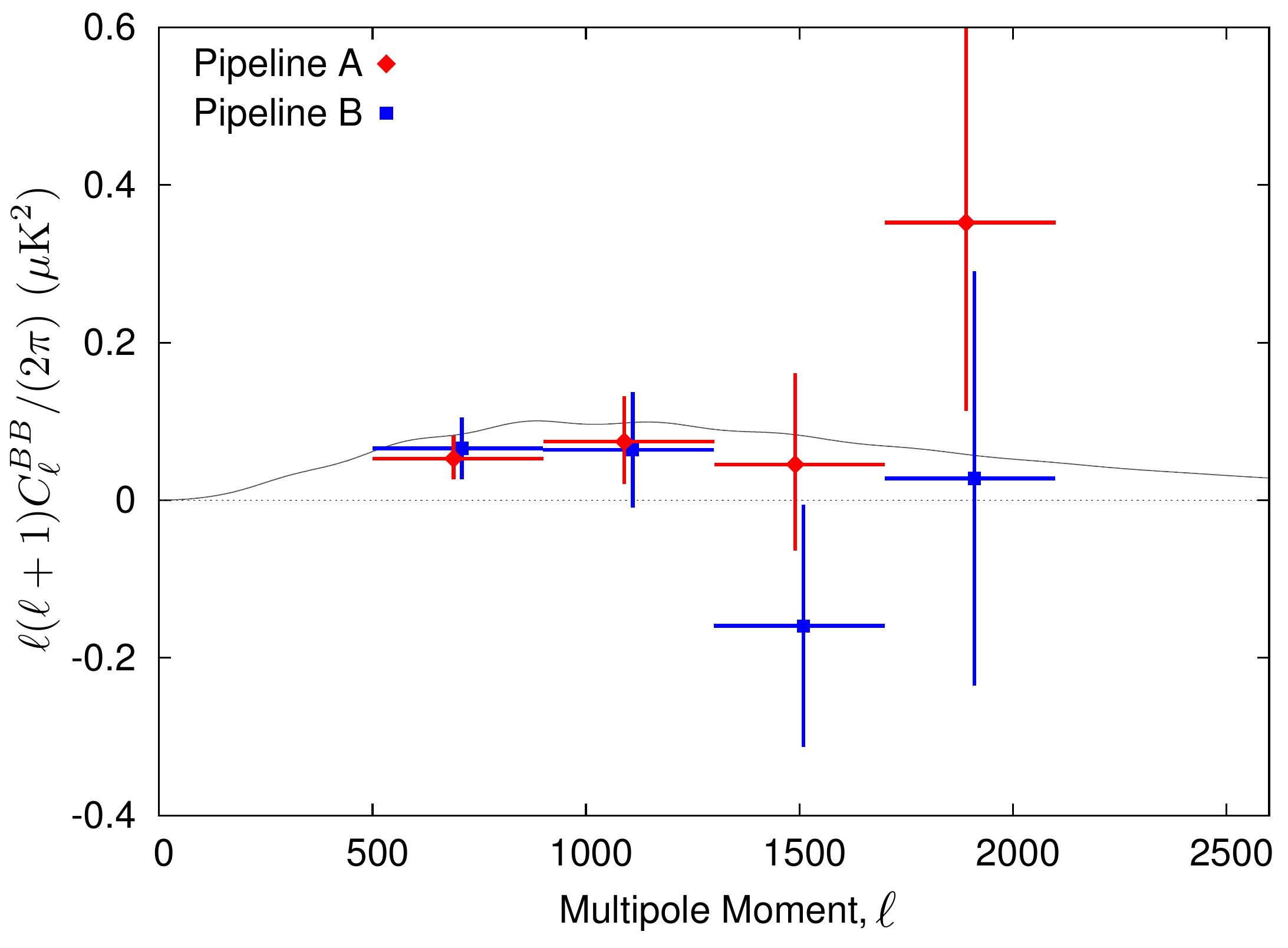}
 \caption{\label{fig:resultBB}\textsc{Polarbear}{} $B$-mode angular power spectrum from the two-season datasets.
 Red diamonds~(blue squares) show the measured band powers from pipeline~A~(pipeline~B).
 The plotted error bars correspond to the 68.3\% confidence intervals of the statistical uncertainty only.
 The multiplicative uncertainty~(due to e.g., the absolute calibration uncertainty), and systematic uncertainties are not plotted,
 but can be found in Table~\ref{tab:summary_systematic_uncertainties}.
       The black curve is a theoretical Planck~2015\ $\Lambda$CDM\ lensing $B$-mode spectrum shown for comparison.}
\end{figure*}

The final analysis procedure follows closely the final first season analysis described in PB14.
For each patch, and each spectrum, we first form the covariance matrices using our set of 500~MC simulations (100~MC for pipeline~B), and we then add the beam and gain calibration uncertainties as described later in this section.
We note that the use of MC simulations to construct the band power covariance matrices gives more realistic estimates than the analytical expressions used in PB14.
Finally for both pipelines, a single estimate of the power spectra from the three patches is created using the measured band powers and their covariance matrices.

The band power covariance matrix accounts for sample variance and instrumental noise variance.
These contributions are estimated by means of 500~MC simulations which use the same pointing and detector weighting as the data, and take into account signal from a beam-convolved realization of a Planck~2015{} $\Lambda$CDM\ power spectrum that includes the effect of gravitational lensing. 
As in PB14, we model the noise in the simulation as white, and we add random spectrally flat noise to the TOD of each detector variance equivalent to that measured from the detectors.
For each detector, the noise variance is estimated from the average of the time domain power spectral densities over 1--3\,Hz, corresponding approximately to an $\ell$ range of 500--1500, and the filtering of the TOD is designed to suppress the correlated part of the noise in this frequency range.
This choice of using white noise, as opposed to a model of correlated noise, has been validated with the null-test framework~(Sec.~\ref{sec:sys-null}).

Finally, we combine the per-patch power spectra into a single patch-combined spectrum according to the band power covariance matrices.
The diagonal elements of the patch-combined covariance matrix are used as the pipeline-B{} error bars shown in Figure~\ref{fig:resultBB}.
Note that the multiplicative uncertainty~(due to e.g., the absolute calibration uncertainty), and systematic uncertainties are not shown in Figure~\ref{fig:resultBB},
but can be found in Table~\ref{tab:summary_systematic_uncertainties}.
For the released $B$-mode band powers from pipeline~A, we go a step further and account for the slightly non-Gaussian shape of the band power likelihoods using an offset log-normal distribution \citep{0004-637X-533-1-19}:
\begin{align}
-2\ln {\cal{L}}(X_{\rm meas} | X_{\rm cosm}) =& \dfrac{1}{\sigma^2} \Big[ \ln[X_{\rm meas} + x_0] \nonumber \\
& - \ln[X_{\rm cosm} + x_0] \Big]^2  \nonumber \\
&- \ln[\sigma (X_{\rm meas} + x_0)\sqrt{2\pi}],
\label{eq:likelihood}
\end{align}
where $X_{\rm meas}$ is the measure of a parameter of interest $X$, and $X_{\rm cosm}$ a fiducial value for this parameter (the $B$-mode band powers in this case).
We fit for the values of $\sigma$ and $x_0$ in each band using simulations with 0\%, 69\% and 100\% of the $B$-mode power predicted by the Planck~2015\ $\Lambda$CDM\ model. 
The intermediate value of 69\% was chosen as being close to the observed $B$-mode power. 
The values are reported in Table~\ref{tab:bandpowers} along with the band powers.
We use the same formalism to empirically model the likelihood for the $B$-mode amplitude parameters $A_{BB}$ and $A_L$ described below.
{\bf We find the average of the upper and lower error bar of the band powers~(the $B$-mode amplitude) is few $\%$ smaller or larger ($\sim 7\%$ smaller) than
that from using the Gaussian distribution. The shape of the likelihood is positively skewed and the upper error bar is $\lesssim 10\%$~($\sim 7\%$) larger than the lower error bar.}

The combined $C_\ell^{TT}$, $C_\ell^{EE}$, $C_\ell^{TE}$, $C_\ell^{TB}$, and $C_\ell^{EB}$\ spectra after the global calibration of absolute gain and polarization angle are plotted in Figure~\ref{fig:spectrumresults}.
We find that the patch-combined and individual patch spectra are consistent with the $\Lambda$CDM\ model.
The patch-combined spectra from pipeline~A~(pipeline~B) have a PTE with respect to the $\Lambda$CDM{} cosmology of 32\%~(28\%), 62\%(26\%), 92\%(67\%), 79\%(29\%), and 60\%(75\%) for $C_\ell^{TT}$, $C_\ell^{EE}$, $C_\ell^{TE}$, $C_\ell^{TB}$, and $C_\ell^{EB}$, respectively.
The spectra from both pipelines are in good agreement.

The difference of the individual-patch $C_\ell^{BB}$\ spectra from pipeline~A\ and pipeline~B\
has a PTE with respect to a null spectrum of 23\%, where the variance of the null spectrum is derived from an analytical estimate.
We remind the reader that the two pipelines treat noisy modes differently (see~\cite{explicit_map_making} for a detailed discussion), and we therefore expect a larger scatter between results in the noise dominated regime ($\ell > 1300$).

The combined $C_\ell^{BB}$\ spectra are shown in Figure~\ref{fig:resultBB}, indicating that the spectra from both pipelines are in good agreement.
The PTE of these band powers with respect to the Planck~2015{} $\Lambda$CDM\ spectrum are 55\%~(pipeline~A) and 41\%~(pipeline~B).
Since the results using pipeline~A{} has satisfied a larger set of the null tests, and the instrumental systematic error analysis was fully performed only on pipeline~A, we adopt its power spectra results~(tabulated in Table~\ref{tab:bandpowers}) as the reference~(official) \textsc{Polarbear}{} results for this release.
The results in Table~\ref{tab:bandpowers} are the data that should be quoted or used for any further cosmological analysis.
\begin{table}
\begin{center}
\caption{\label{tab:bandpowers}Reported \textsc{Polarbear}\ band powers $D_\ell^{BB}$$=$$\ell \left ( \ell + 1 \right ) C_\ell^{BB} / 2\pi$\ from pipeline~A}
\begin{tabular}{c|c}
\hline
\hline
Central~$\ell$ & $D_\ell^{BB}$~[$\mu$K$^2$] \\
\hline
\phantom{0}700 & $0.053 ^{+0.029} _{-0.026}$ \\
          1100 & $0.074 ^{+0.057} _{-0.054}$ \\
          1500 & $0.045 ^{+0.116} _{-0.109}$ \\
          1900 & $0.352 ^{+0.260} _{-0.239}$ \\
\hline
\end{tabular}
 \tablecomments{The errors correspond to the 68.3\% confidence intervals of the statistical uncertainty only.
 The multiplicative uncertainty~(due to e.g., the absolute calibration uncertainty), and systematic uncertainties are not included,
 but can be found in Table~\ref{tab:summary_systematic_uncertainties}.
   Correlations between neighboring bins are small and consistent with zero
 within the statistical uncertainty of $\pm 0.05$ due to the finite number of Monte Carlo-simulation realizations.}
\end{center}
\end{table}

First we consider the significance with which these data rule out the null hypothesis of no $B$-modes. 
After setting the sample variance to zero, we fit for the amplitude parameter, finding $A_{BB} = 0.75 ^{+0.21} _{-0.20} ({\rm stat}) ^{+0.00}_{-0.04}({\rm inst})$, where in this expression,
``stat'' refers to the expected statistical fluctuation of the measurement~($X_{\rm meas}$ in Equation~\ref{eq:likelihood} with $X=A_{BB}$)
evaluated with the likelihood of no lensing signal~($X_{\rm cosm}=0$), and ``inst'' to the systematic uncertainty associated with possible biases from the instrument.
To calculate the upper bound on the additive uncertainties from instrumental systematic errors, we linearly add, in each band, the upper bound of the band powers of all the instrument systematic effects discussed in Sec.~\ref{sec:systematic_pipeline}.
This produces a lower $A_{BB}$, and sets the lower bound of the additive uncertainty.
To be conservative, we evaluate the detection significance by subtracting this systematic uncertainty of 0.04 from 0.75.
The likelihood ratio is ${\cal{L}}(X_{\rm meas}=0.71|X_{\rm cosm}=0) / {\rm max}_{X_{\rm cosm}} {\cal{L}}(X_{\rm meas}=0.71|X_{\rm cosm}) = 9.0 \times 10^{-3}$,
corresponding to a 3.1$\sigma${} rejection of the no $B$-mode null hypothesis.
\begin{table*}[htbp]
\begin{center}
 \caption{\label{tab:summary_systematic_uncertainties}Summary of the reported \textsc{Polarbear}\ systematic uncertainties.}
\begin{tabular}{r p{2.6in} p{0.85in} p{0.79in}}
\hline
\hline
\centering Type &\centering Source of systematics & \centering Effect on $D_\ell^{BB}$\ [$10^{-4}$\,$\mu$K$^2$] & Effect on $A_{BB}$ \\
 \hline
 Instrument~(Sec.~\ref{sec:systematic_pipeline})
 &	Gain drift         & \hspace{11pt}$\phantom{0}8.5$ & \hspace{14pt}$0.009$\\  
 &	Differential gain  & \hspace{11pt}$\phantom{0}9.3$ & \hspace{14pt}$0.010$\\ 
 & Differential beam size  & \hspace{11pt}$\phantom{0}0.4$ & \hspace{14pt}$0.000$\\ 
 & Differential beam ellipticity  & \hspace{11pt}$\phantom{0}0.1$ & \hspace{14pt}$0.000$\\ 
 & Differential \& Boresight pointing  & \hspace{11pt}$\phantom{0}5.7$ & \hspace{14pt}$0.008$\\ 
 & Instrument \& Relative polarization angle  & \hspace{11pt}$\phantom{0}6.7$ & \hspace{14pt}$0.008$\\ 
 & Electrical crosstalk  & \hspace{11pt}$\phantom{0}2.5$ & \hspace{14pt}$0.003$\\ 
 \cline{2-4}
 & Total  & \hspace{11pt}$33.3$ & \hspace{14pt}$0.037$\\ 
\hline
 Astrophysical foreground
 & Galactic dust~(Sec.~\ref{sec:sys-fg-diffuse})  & $63.5 \pm 123.3$ & $0.071 \pm 0.138$ \\
 & Galactic synchrotron~(Sec.~\ref{sec:sys-fg-diffuse}) & $\phantom{0}1.4 \pm \phantom{00}2.1$ & $0.002 \pm 0.002$\\
 & Radio \& Dusty galaxies~(Sec.~\ref{sec:sys-fg-radio_and_dusty_galaxies}) & $13.4 \pm \phantom{00}5.5$ & $0.019 \pm 0.005$\\
 \cline{2-4}
 & sub total & $78.3 \pm 123.4$ & $0.092 \pm 0.138$\\
 \cline{2-4}
 Analysis\footnote{The residual of the ground pickup removal and $E$-to-$B$ leakage subtraction
 and the sensitivity of the transfer function to different cosmologies are evaluated using the same methodology of PB14.
 The systematic bias and uncertainty from ``Analysis'' is subtracted from the measured band powers and propagated in pipeline~A.}
 &      Ground pickup removal & $\phantom{0}0.5 \pm \phantom{00}1.7$ & $0.001 \pm 0.002$\\
 & $E$-to-$B$ leakage due to filter subtraction & $\phantom{0}2.5$& $0.003$\\
 \cline{2-4}
 & sub total  & $\phantom{0}2.9 \pm \phantom{00}1.7$ & $0.003 \pm 0.002$\\
 \cline{2-4}
& Total & $81.2 \pm 123.4$ & $0.095 \pm 0.138$\\
\tableline
\tableline
 Multiplicative effect
 & Absolute gain uncertainty~(Sec.~\ref{sec:absgain_and_polangle}) & & $\phantom{0.000}\pm 3.0\%$ \\ 
 & Beam uncertainty~(Sec.~\ref{sec:absgain_and_polangle}) & & $\phantom{0.000}\pm 1.0\%$ \\ 
 & Polarization efficiency~(Sec.~\ref{sec:polarization}) & & $\phantom{0.000}\pm 3.3\%$ \\
 & Transfer function\tablenotemark{a} & & $\phantom{0.000}\pm 3.9\%$ \\
 \cline{2-4}
 & Total & & $\phantom{0.000}\pm 6.0\%$\\
 \hline\hline
\end{tabular}
\tablecomments{Note the third column represents effect on $D_\ell^{BB}${} at the first $\ell=[500, 900]$ bin,
 on the other hand, the fourth column represents total effect on $A_L${} in the multipole range $500 \leq \ell \leq 2100$.}
\end{center}
\end{table*}

\begin{figure}[htbp]
 \centering
 \includegraphics[width=3.3in]{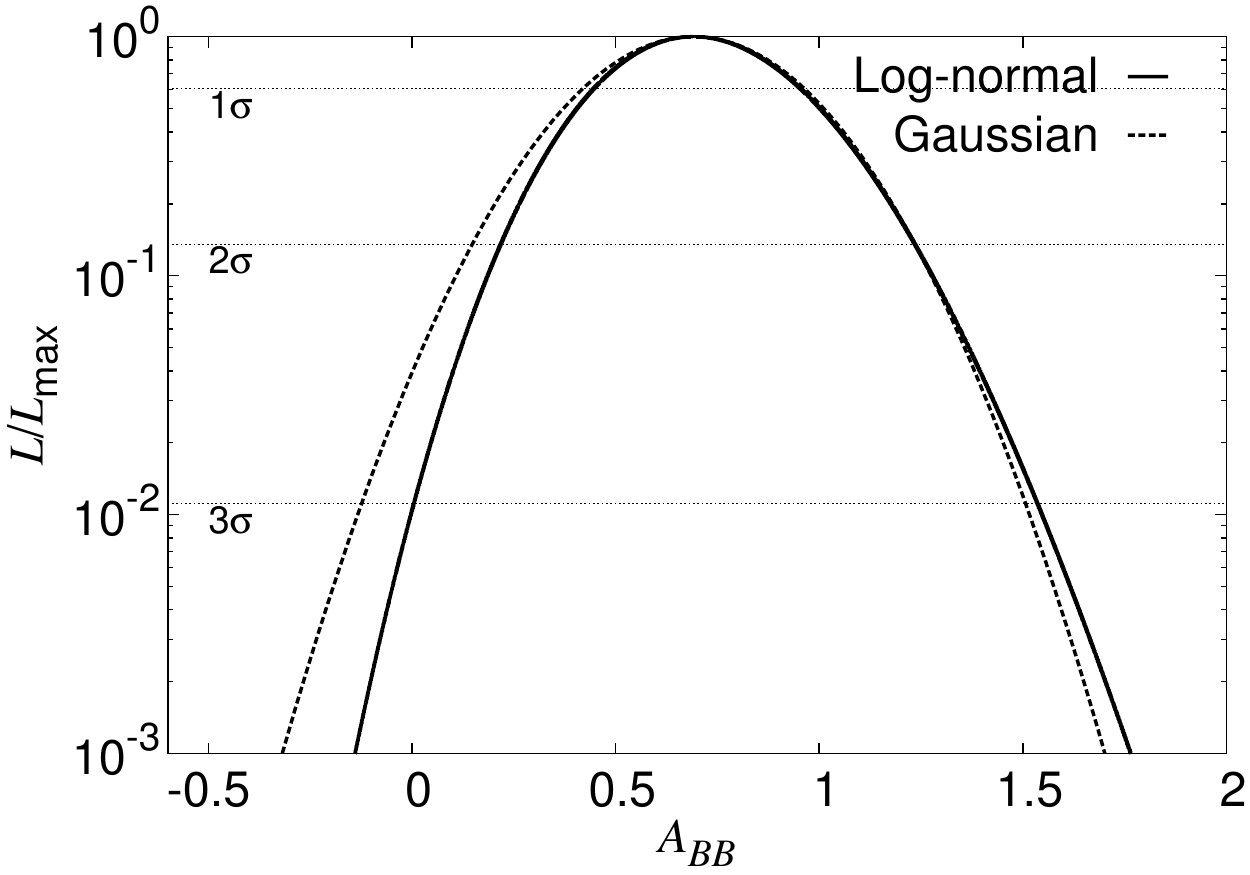}
 \caption{\label{fig:likelihood_curve_of_ABB}
 Likelihood curve of the $B$-mode amplitde modeled by an offset log-normal distribution is shown.
 That by the Gaussian distribution is also shown for comparison.
  The intersections of the curves with the dashed horizontal lines give the bounds with $1 \sigma$, $2 \sigma$, and $3 \sigma$ confidence intervals.
 The log-normal functional form captures the asymmetric shape of the likelihood function due to sample variance.}
\end{figure}

Next we fit the band powers to the Planck~2015{} $\Lambda$CDM\ cosmological model with a single $B$-mode amplitude parameter, $A_{BB}$. 
We find $A_{BB} = 0.69 ^{+0.26} _{-0.25} ({\rm stat}) ^{+0.00}_{-0.04}({\rm inst}) \pm 0.04 ({\rm multi})$, where in this expression,
``stat'' refers to the 68.3\% confidence interval of the estimated quantity~($X_{\rm cosm}$ in Equation~\ref{eq:likelihood} with $X=A_{BB}$)
given our observation with non-zero lensing signal~($X_{\rm meas}=0.69$), and ``multi'' to multiplicative calibration uncertainties.
{\bf The likelihood of the $B$-mode amplitude is shown in Figure~\ref{fig:likelihood_curve_of_ABB}.}
The shift in the estimated amplitude with respect to the null result is due to different field and bandpower weighting,
which uses the covariance matrix with the Planck~2015{} $\Lambda$CDM{} model, or $A_{BB}=1$.

Finally, we fit for the amplitude parameter $A_{BB}$\ after subtracting the foreground terms from Sec.~\ref{sec:sys-fg}, which we denote finally as $A_L$. 
We find $A_L = 0.60 ^{+0.26} _{-0.24} ({\rm stat}) ^{+0.00} _{-0.04}({\rm inst}) \pm 0.14 ({\rm foreground}) \pm 0.04 ({\rm multi})$, where ``foreground'' refers to the total foreground uncertainty.
This amplitude can be interpreted as the measured amplitude of the lensing $B$-modes. 
Table~\ref{tab:summary_systematic_uncertainties} summarizes all the systematic uncertainties in the measurement of $A_L$.
Although the mean value shifts from the previous measurement in PB14, the change in the mean value is consistent with statistical fluctuation even accounting for the fact that data have overlap between this result and PB14.
Compared to PB14, the uncertainty on the lensing amplitude has been reduced by a factor of two thanks to the realistic uncertainty estimate as well as the overall larger dataset and improved calibration.

\begin{figure*}[htbp]
 \centering
 \includegraphics[width=5.5in]{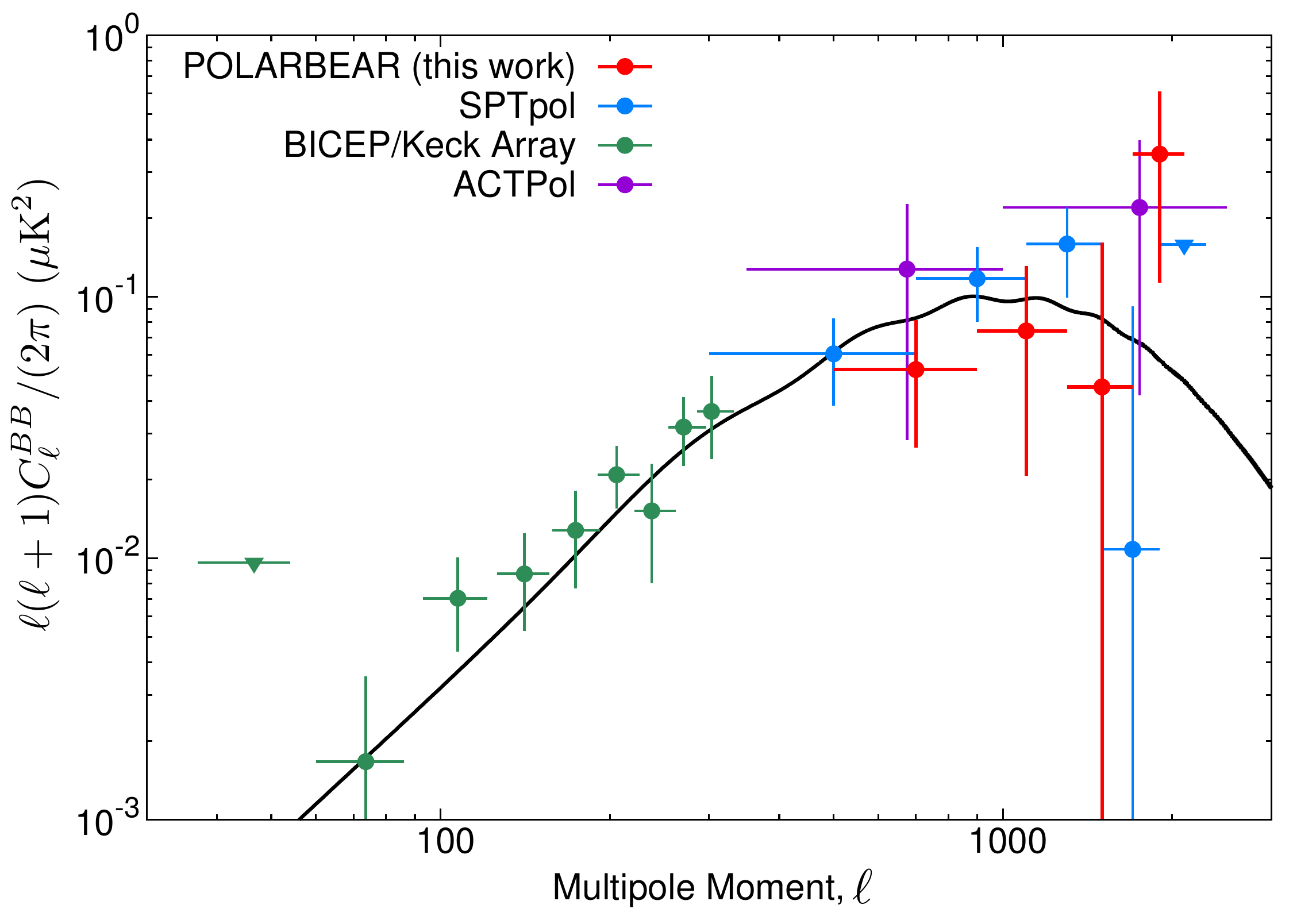}
 \caption{\label{fig:resultBB_w_others}
 $B$-mode polarization power spectrum measurements from \textsc{Polarbear}~(this work by pipeline~A), {\sc SPTpol}~\citep{Keisler2015},
 \textsc{actpol}~\citep{2016arXiv161002360L}, \textit{Keck Array}~\citep{bkV}. Uncertainties correspond to a 68.3\% confidence level, while upper limits are quoted at a 95.4\% confidence level. The black curve is a theoretical Planck~2015{} $\Lambda$CDM\ spectrum.}
\end{figure*}
 
\section{Conclusions}
We have measured the CMB $B$-mode angular power spectrum, $C_\ell^{BB}$, over the multipole range $500 < \ell < 2100$ from a blind analysis of data from the first two seasons of \textsc{Polarbear}\ observations.
We have doubled the sensitivity on the lensing amplitude compared to the first-season result (PB14) and rejected the null hypothesis of no $B$-mode polarization with 3.1$\sigma${} confidence. {\bf After subtracting the new estimated foreground contamination based on the Planck~2015{} data which were unavailable in PB14,}
we found the amplitude of the lensing signal to be $A_L = 0.60 ^{+0.26} _{-0.24} ({\rm stat}) ^{+0.00} _{-0.04}({\rm inst}) \pm 0.14 ({\rm foreground}) \pm 0.04 ({\rm multi})$.

The data were analyzed by two independent pipelines giving results in agreement.
These results are supported by an extensive suite of null tests in which 12 divisions of data were used to finalize the data selection and by an estimate of systematic errors from 9 sources of instrumental contamination using a detailed instrument model.
We found that all the systematic uncertainties were small compared to the statistical uncertainty in the measurement. 
To motivate comprehensive evaluation of the dataset and prevent observer bias in data selection and analysis,
the selection criteria and systematic errors were determined before the $B$-mode power spectra themselves were examined by the two independent analysis pipelines.
In Figure~\ref{fig:resultBB_w_others}, we present our measurement alongside recent measurements by {\sc SPTpol}~\citep{Keisler2015},
\textsc{actpol}~\citep{2016arXiv161002360L} and \textit{Keck Array}~\citep{bkV}.

We reported previously evidence for gravitational lensing of the polarized CMB in \textsc{Polarbear}\ data~\citep{pb2014b} on the same sky area as studied in this work using the PB14 dataset.
An updated lensing analysis using the first two seasons of \textsc{Polarbear}{} observations described in this paper will be presented soon in a separate publication.
After the two seasons reported here, we installed a continuously rotating half-wave plate in May 2014~\citep{1475-7516-2017-05-008}, and we are observing a larger patch of the sky ($\sim700$ $\deg^2$) since then, targeting inflationary $B$ modes at degree angular scales.
Results from the analysis of this dataset will be reported in future publications.

Foreground contamination is one of the main limiting factors in our measurement of $B$ modes.
Neither {\sc wmap}\ nor Planck~2015\ has enough sensitivity at small angular scales to sufficiently constrain the polarized synchrotron and dust amplitude in the \textsc{Polarbear}\ observations reported here. Dedicated multi-frequency observations are therefore needed to obtain better constraints.
New receivers~(\textsc{Polarbear}-2) at $95\,{\rm GHz}$, $150\,{\rm GHz}$, $220\,{\rm GHz}$ and $270\,{\rm GHz}$ with the sensitivity
to reach inflationary tensor-to-scalar ratio $\sigma(r)<0.01$
are under development~\citep{2016SPIE.9914E..1II} and are being implemented on the Simons Array{} telescopes~\citep{2016JLTP..184..805S}.
Such improvements will enable the Simons Array{} to enter into an era of precise CMB polarization measurements, improving our understanding of the early and late Universe physics.

\acknowledgements
JP acknowledges support from the Science and Technology Facilities Council [grant number ST/L000652/1] and from the European Research Council under the European Union's Seventh Framework Programme (FP/2007-2013) / ERC Grant Agreement No. [616170].
The \textsc{Polarbear}{} project is funded by the National Science Foundation under Grants No. AST-0618398 and No. AST-1212230.
The James Ax Observatory operates in the Parque Astron\'omico Atacama in Northern Chile under the auspices of the Comisi\'on Nacional de Investigaci\'on Cient\'ifica y Tecnol\'ogica de Chile (CONICYT).
This research used resources of the Central Computing System, owned and operated by the Computing Research Center at KEK, the HPCI system (Project ID:hp150132), and the National Energy Research Scientific Computing Center, a DOE Office of Science User Facility supported by the Office of Science of the U.S. Department of Energy under Contract No.  DE-AC02-05CH11231.
In Japan, this work was supported by MEXT KAKENHI Grant Numbers JP15H05891, 21111002, JSPS KAKENHI Grant Numbers JP26220709, JP24111715, JP26800125.
This work was supported by World Premier International Research Center Initiative (WPI), MEXT, Japan.
This work was supported by JSPS Core-to-Core program.
In Italy, this work was supported by the RADIOFOREGROUNDS grant of the European Union's Horizon 2020 research and innovation programme (COMPET-05-2015, grant agreement number 687312) as well as by the INDARK INFN Initiative.
The McGill authors acknowledge funding from the Natural Sciences and Engineering Research Council of Canada and the Canadian Institute for Advanced Research.
GF acknowledges support from the CNES postdoctoral program.
DB acknowledges support from NSF grant AST-1501422.
DB acknowledges support from FONDECYT postdoctoral grant number 3150504.
CF acknowledges support from grants HST-AR-13886.001-A, IGPP LANL 368641, NASA NNX16AF39G and Ax Foundation for Cosmology at UC Irvine.
CR acknowledges support from a Australian Research Council's Future Fellowship (FT150100074).
MA acknowledges support from CONICYT UC Berkeley-Chile Seed Grant (CLAS fund) Number 77047, Fondecyt project 1130777, DFI postgraduate scholarship program and DFI Postgraduate Competitive Fund for Support in the Attendance to Scientific Events.
RD acknowledges support from CONICYT grants FONDECYT 1141113, QUIMAL 160009 and Anillo ACT-1417.
ST was supported by Grant-in-Aid for JSPS Research Fellow.

\begin{appendix}
\section{Detector gains: modeling Saturn's rings} \label{app:saturn_rings}
For the PB14 data release, we used only observations of Saturn taken between June and September 2012 (4 months of data), and we assumed the temperature of Saturn to be constant, with T$_{\text{sat}}=148$ K$_{\text{RJ}}$.
However, the analysis of Saturn observations over a longer period (May 2012 -- April 2014) exhibited variations up to 10\% over time \citep{peloton:tel-01257383}.
We investigated several possible explanations, such as a misestimation of the apparent diameter of the planet, dependency in the elevation, miscorrection of the atmosphere contribution, stability of the stimulator, change in the main beam parameters, or even changes in the sidelobes over time, but none gave an explanation for the shift of approximately $10\%$ in between the two seasons.
This shift was however explained by taking into account the influence of Saturn's rings on the planet flux.
Even if we cannot resolve the rings,\footnote{Given the angular size of Saturn with respect to our beam size, we cannot resolve the details of the planet. The variations in the atmosphere of Saturn (pole/equator differences, clouds, seasonal variations, etc.) are also not taken into account here.} their inclination with respect to the plane of observation from the Earth produces a variation over time in the measured microwave brightness temperature, as described and shown in, e.g., \cite{dunn2002more}, \cite{weiland2011seven}, and \cite{hasselfield2013atacama}.
The contribution of the rings to the total temperature brightness of Saturn is twofold: they obscure and therefore reduce the flux from the main disk, but they also radiate at a lower temperature and contribute to the total signal through a mix of scattering and thermal re-emission of planetary radiation (at 150 GHz, the thermal emission dominates over the scattering).
The change in opening angle (i.e., the angle made by the plane of the rings and the line of sight of the observer) allows us to break the degeneracy between the contribution of the disk and that of the rings.
We follow the two-parameter empirical model proposed by {\sc wmap}\ \citep{weiland2011seven}, by modeling the total brightness temperature of Saturn as a contribution from the disk $T_{\text{disk}}$ with seven extended surrounding rings described by one global effective temperature $T_{\text{ring}}$.\footnote{$T_{\text{disk}}$ and $T_{\text{ring}}$ temperatures are assumed time invariant.}
The individual contributions from those two parameters to the total brightness is given by the orientation of the system at the moment of the observation.

We use 132 observations of Saturn between May 2012 and April 2014, selected for their high quality of data, and spanning ring opening angles from $\sim12^{\circ}$ to $\sim23^{\circ}$.
The observations are combined into periods of 15--30 days, taking the uncertainty in each combined measurement to be the error in the mean of the contributing data.
The model fit parameters and uncertainties ($\chi^2\sim2$ per degree of freedom) are $T_{\text{disk}} = 126.8 \pm 2.7$ K$_{\text{RJ}}$\ and $T_{\text{ring}} = 16.3 \pm 6.0$ K$_{\text{RJ}}$\ (after applying the absolute calibration derived from the CMB temperature spectrum as described in Sec.~\ref{sec:absgain_and_polangle}).
The uncertainties include a statistical contribution and a systematic contribution from the wafer-to-wafer variations.
{\sc act}\ collaboration provides complementary measurements at 148\,GHz for ring opening angles from $-2^{\circ}$ to $12^{\circ}$ in \cite{hasselfield2013atacama}.
They found $T_{\text{disk}} = 133.8 \pm 3.2$ K$_{\text{RJ}}$\ and $T_{\text{ring}} = 17.7 \pm 2.2$ K$_{\text{RJ}}$, which is consistent with our values.
On the other hand Planck~2015\ collaboration reported recently a measurement of Saturn's disk temperature rescaled at 147 GHz~\citep{2016arXiv161207151P}. They found a temperature of the main disk $T_{\text{disk}} = 145.7 \pm 1.1$ K$_{\text{RJ}}$\ higher than the \textsc{Polarbear}\ and {\sc act}\ values. However, Figure~10 in~\cite{2016arXiv161207151P} seems to suggest that the ring temperature between Planck~2015\ and {\sc act}, and therefore \textsc{Polarbear}, are in agreement. \end{appendix}

\bibliographystyle{apj}

\bibliography{spp}

\newcommand{\noopsort}[1]{}
\begin{thebibliography}{}
\expandafter\ifx\csname natexlab\endcsname\relax\def\natexlab#1{#1}\fi

\bibitem[{{Arnold} {et~al.}(2012){Arnold}, {Ade}, {Anthony}, {Barron},
  {Boettger}, {Borrill}, {Chapman}, {Chinone}, {Dobbs}, {Errard}, {Fabbian},
  {Flanigan}, {Fuller}, {Ghribi}, {Grainger}, {Halverson}, {Hasegawa},
  {Hattori}, {Hazumi}, {Holzapfel}, {Howard}, {Hyland}, {Jaffe}, {Keating},
  {Kermish}, {Kisner}, {Le Jeune}, {Lee}, {Linder}, {Lungu}, {Matsuda},
  {Matsumura}, {Miller}, {Meng}, {Morii}, {Moyerman}, {Myers}, {Nishino},
  {Paar}, {Quealy}, {Reichardt}, {Richards}, {Ross}, {Shimizu}, {Shimmin},
  {Shimon}, {Sholl}, {Siritanasak}, {Spieler}, {Stebor}, {Steinbach},
  {Stompor}, {Suzuki}, {Tomaru}, {Tucker}, \& {Zahn}}]{Arnold_SPIE2012}
{Arnold}, K., {Ade}, P.~A.~R., {Anthony}, A.~E., {et~al.} 2012, in Society of
  Photo-Optical Instrumentation Engineers (SPIE) Conference Series, Vol. 8452,
  Society of Photo-Optical Instrumentation Engineers (SPIE) Conference Series

\bibitem[{{Aumont} {et~al.}(2010){Aumont}, {Conversi}, {Thum}, {Wiesemeyer},
  {Falgarone}, {Mac{\'{\i}}as-P{\'e}rez}, {Piacentini}, {Pointecouteau},
  {Ponthieu}, {Puget}, {Rosset}, {Tauber}, \& {Tristram}}]{2010A&A...514A..70A}
{Aumont}, J., {Conversi}, L., {Thum}, C., {et~al.} 2010, \aap, 514, A70

\bibitem[{{Bennett} {et~al.}(2013){Bennett}, {Larson}, {Weiland}, {Jarosik},
  {Hinshaw}, {Odegard}, {Smith}, {Hill}, {Gold}, {Halpern}, {Komatsu}, {Nolta},
  {Page}, {Spergel}, {Wollack}, {Dunkley}, {Kogut}, {Limon}, {Meyer}, {Tucker},
  \& {Wright}}]{2013ApJS..208...20B}
{Bennett}, C.~L., {Larson}, D., {Weiland}, J.~L., {et~al.} 2013, \apjs, 208, 20

\bibitem[{{Bonavera} {et~al.}(2017){Bonavera}, {Gonz{\'a}lez-Nuevo},
  {Arg{\"u}eso}, \& {Toffolatti}}]{bonavera17}
{Bonavera}, L., {Gonz{\'a}lez-Nuevo}, J., {Arg{\"u}eso}, F., \& {Toffolatti},
  L. 2017, ArXiv e-prints, arXiv:1703.09952

\bibitem[{Bond {et~al.}(2000)Bond, Jaffe, \& Knox}]{0004-637X-533-1-19}
Bond, J.~R., Jaffe, A.~H., \& Knox, L. 2000, The Astrophysical Journal, 533, 19

\bibitem[{{De Zotti} {et~al.}(2005){De Zotti}, {Ricci}, {Mesa}, {Silva},
  {Mazzotta}, {Toffolatti}, \& {Gonz{\'a}lez-Nuevo}}]{dezotti}
{De Zotti}, G., {Ricci}, R., {Mesa}, D., {et~al.} 2005, \aap, 431, 893

\bibitem[{Dunn {et~al.}(2002)Dunn, Molnar, \& Fix}]{dunn2002more}
Dunn, D.~E., Molnar, L.~A., \& Fix, J.~D. 2002, Icarus, 160, 132

\bibitem[{{Fert{\'e}} {et~al.}(2013){Fert{\'e}}, {Grain}, {Tristram}, \&
  {Stompor}}]{ferte2013}
{Fert{\'e}}, A., {Grain}, J., {Tristram}, M., \& {Stompor}, R. 2013, \prd, 88,
  023524

\bibitem[{{Fuskeland} {et~al.}(2014){Fuskeland}, {Wehus}, {Eriksen}, \&
  {N{\ae}ss}}]{Fuskeland14}
{Fuskeland}, U., {Wehus}, I.~K., {Eriksen}, H.~K., \& {N{\ae}ss}, S.~K. 2014,
  ApJ, 790, 104

\bibitem[{{George} {et~al.}(2015){George}, {Reichardt}, {Aird}, {Benson},
  {Bleem}, {Carlstrom}, {Chang}, {Cho}, {Crawford}, {Crites}, {de Haan},
  {Dobbs}, {Dudley}, {Halverson}, {Harrington}, {Holder}, {Holzapfel}, {Hou},
  {Hrubes}, {Keisler}, {Knox}, {Lee}, {Leitch}, {Lueker}, {Luong-Van},
  {McMahon}, {Mehl}, {Meyer}, {Millea}, {Mocanu}, {Mohr}, {Montroy}, {Padin},
  {Plagge}, {Pryke}, {Ruhl}, {Schaffer}, {Shaw}, {Shirokoff}, {Spieler},
  {Staniszewski}, {Stark}, {Story}, {van Engelen}, {Vanderlinde}, {Vieira},
  {Williamson}, \& {Zahn}}]{2015ApJ...799..177G}
{George}, E.~M., {Reichardt}, C.~L., {Aird}, K.~A., {et~al.} 2015, \apj, 799,
  177

\bibitem[{{G{\'o}rski} {et~al.}(2005){G{\'o}rski}, {Hivon}, {Banday},
  {Wandelt}, {Hansen}, {Reinecke}, \& {Bartelmann}}]{Gorski2005}
{G{\'o}rski}, K.~M., {Hivon}, E., {Banday}, A.~J., {et~al.} 2005, \apj, 622,
  759

\bibitem[{Grain {et~al.}(2009)Grain, Tristram, \& Stompor}]{xpure}
Grain, J., Tristram, M., \& Stompor, R. 2009, Phys. Rev. D, 79, 123515

\bibitem[{{Grain} {et~al.}(2012){Grain}, {Tristram}, \& {Stompor}}]{Grain2012}
{Grain}, J., {Tristram}, M., \& {Stompor}, R. 2012, \prd, 86, 076005

\bibitem[{Hasselfield {et~al.}(2013)Hasselfield, Moodley, Bond, Das, Devlin,
  Dunkley, D{\"u}nner, Fowler, Gallardo, Gralla,
  {et~al.}}]{hasselfield2013atacama}
Hasselfield, M., Moodley, K., Bond, J.~R., {et~al.} 2013, The Astrophysical
  Journal Supplement Series, 209, 17

\bibitem[{{Hivon} {et~al.}(2002){Hivon}, {G{\'o}rski}, {Netterfield}, {Crill},
  {Prunet}, \& {Hansen}}]{Hivon2002}
{Hivon}, E., {G{\'o}rski}, K.~M., {Netterfield}, C.~B., {et~al.} 2002, \apj,
  567, 2

\bibitem[{{Inoue} {et~al.}(2016){Inoue}, {Ade}, {Akiba}, {Aleman}, {Arnold},
  {Baccigalupi}, {Barch}, {Barron}, {Bender}, {Boettger}, {Borrill}, {Chapman},
  {Chinone}, {Cukierman}, {de Haan}, {Dobbs}, {Ducout}, {D{\"u}nner},
  {Elleflot}, {Errard}, {Fabbian}, {Feeney}, {Feng}, {Fuller}, {Gilbert},
  {Goeckner-Wald}, {Groh}, {Hall}, {Halverson}, {Hamada}, {Hasegawa},
  {Hattori}, {Hazumi}, {Hill}, {Holzapfel}, {Hori}, {Howe}, {Irie}, {Jaehnig},
  {Jaffe}, {Jeong}, {Katayama}, {Kaufman}, {Kazemzadeh}, {Keating}, {Kermish},
  {Keskitalo}, {Kisner}, {Kusaka}, {Le Jeune}, {Lee}, {Leon}, {Linder},
  {Lowry}, {Matsuda}, {Matsumura}, {Miller}, {Mizukami}, {Montgomery},
  {Navaroli}, {Nishino}, {Paar}, {Peloton}, {Poletti}, {Puglisi}, {Raum},
  {Rebeiz}, {Reichardt}, {Richards}, {Ross}, {Rotermund}, {Segawa}, {Sherwin},
  {Shirley}, {Siritanasak}, {Stebor}, {Stompor}, {Suzuki}, {Suzuki}, {Tajima},
  {Takada}, {Takatori}, {Teply}, {Tikhomirov}, {Tomaru}, {Whitehorn}, {Zahn},
  \& {Zahn}}]{2016SPIE.9914E..1II}
{Inoue}, Y., {Ade}, P., {Akiba}, Y., {et~al.} 2016, in \procspie, Vol. 9914,
  Society of Photo-Optical Instrumentation Engineers (SPIE) Conference Series,
  99141I

\bibitem[{{Keating} {et~al.}(2013){Keating}, {Shimon}, \& {Yadav}}]{ksy2013}
{Keating}, B.~G., {Shimon}, M., \& {Yadav}, A.~P.~S. 2013, \apjl, 762, L23

\bibitem[{{Keisler} {et~al.}(2015){Keisler}, {Hoover}, {Harrington}, {Henning},
  {Ade}, {Aird}, {Austermann}, {Beall}, {Bender}, {Benson}, {Bleem},
  {Carlstrom}, {Chang}, {Chiang}, {Cho}, {Citron}, {Crawford}, {Crites}, {de
  Haan}, {Dobbs}, {Everett}, {Gallicchio}, {Gao}, {George}, {Gilbert},
  {Halverson}, {Hanson}, {Hilton}, {Holder}, {Holzapfel}, {Hou}, {Hrubes},
  {Huang}, {Hubmayr}, {Irwin}, {Knox}, {Lee}, {Leitch}, {Li}, {Luong-Van},
  {Marrone}, {McMahon}, {Mehl}, {Meyer}, {Mocanu}, {Natoli}, {Nibarger},
  {Novosad}, {Padin}, {Pryke}, {Reichardt}, {Ruhl}, {Saliwanchik}, {Sayre},
  {Schaffer}, {Shirokoff}, {Smecher}, {Stark}, {Story}, {Tucker},
  {Vanderlinde}, {Vieira}, {Wang}, {Whitehorn}, {Yefremenko}, \&
  {Zahn}}]{Keisler2015}
{Keisler}, R., {Hoover}, S., {Harrington}, N., {et~al.} 2015, \apj, 807, 151

\bibitem[{{Kermish} {et~al.}(2012){Kermish}, {Ade}, {Anthony}, {Arnold},
  {Barron}, {Boettger}, {Borrill}, {Chapman}, {Chinone}, {Dobbs}, {Errard},
  {Fabbian}, {Flanigan}, {Fuller}, {Ghribi}, {Grainger}, {Halverson},
  {Hasegawa}, {Hattori}, {Hazumi}, {Holzapfel}, {Howard}, {Hyland}, {Jaffe},
  {Keating}, {Kisner}, {Lee}, {Le Jeune}, {Linder}, {Lungu}, {Matsuda},
  {Matsumura}, {Meng}, {Miller}, {Morii}, {Moyerman}, {Myers}, {Nishino},
  {Paar}, {Quealy}, {Reichardt}, {Richards}, {Ross}, {Shimizu}, {Shimon},
  {Shimmin}, {Sholl}, {Siritanasak}, {Spieler}, {Stebor}, {Steinbach},
  {Stompor}, {Suzuki}, {Tomaru}, {Tucker}, \& {Zahn}}]{Kermish_SPIE2012}
{Kermish}, Z.~D., {Ade}, P., {Anthony}, A., {et~al.} 2012, in Society of
  Photo-Optical Instrumentation Engineers (SPIE) Conference Series, Vol. 8452,
  Society of Photo-Optical Instrumentation Engineers (SPIE) Conference Series

\bibitem[{{Krachmalnicoff} {et~al.}(2016){Krachmalnicoff}, {Baccigalupi},
  {Aumont}, {Bersanelli}, \& {Mennella}}]{2016A&A...588A..65K}
{Krachmalnicoff}, N., {Baccigalupi}, C., {Aumont}, J., {Bersanelli}, M., \&
  {Mennella}, A. 2016, A\&A, 588, A65

\bibitem[{{Louis} {et~al.}(2016){Louis}, {Grace}, {Hasselfield}, {Lungu},
  {Maurin}, {Addison}, {Ade}, {Aiola}, {Allison}, {Amiri}, {Angile},
  {Battaglia}, {Beall}, {de Bernardis}, {Bond}, {Britton}, {Calabrese}, {Cho},
  {Choi}, {Coughlin}, {Crichton}, {Crowley}, {Datta}, {Devlin}, {Dicker},
  {Dunkley}, {D{\"u}nner}, {Ferraro}, {Fox}, {Gallardo}, {Gralla}, {Halpern},
  {Henderson}, {Hill}, {Hilton}, {Hilton}, {Hincks}, {Hlozek}, {Ho}, {Huang},
  {Hubmayr}, {Huffenberger}, {Hughes}, {Infante}, {Irwin}, {Muya Kasanda},
  {Klein}, {Koopman}, {Kosowsky}, {Li}, {Madhavacheril}, {Marriage}, {McMahon},
  {Menanteau}, {Moodley}, {Munson}, {Naess}, {Nati}, {Newburgh}, {Nibarger},
  {Niemack}, {Nolta}, {Nu{\~n}ez}, {Page}, {Pappas}, {Partridge}, {Rojas},
  {Schaan}, {Schmitt}, {Sehgal}, {Sherwin}, {Sievers}, {Simon}, {Spergel},
  {Staggs}, {Switzer}, {Thornton}, {Trac}, {Treu}, {Tucker}, {Van Engelen},
  {Ward}, \& {Wollack}}]{2016arXiv161002360L}
{Louis}, T., {Grace}, E., {Hasselfield}, M., {et~al.} 2016, ArXiv e-prints,
  arXiv:1610.02360

\bibitem[{Matsuda(2017)}]{fmatsuda2017}
Matsuda, F. 2017, University of California, San Diego, In Preparation

\bibitem[{{Murphy} {et~al.}(2010){Murphy}, {Sadler}, {Ekers}, {Massardi},
  {Hancock}, {Mahony}, {Ricci}, {Burke-Spolaor}, {Calabretta}, {Chhetri}, {de
  Zotti}, {Edwards}, {Ekers}, {Jackson}, {Kesteven}, {Lindley}, {Newton-McGee},
  {Phillips}, {Roberts}, {Sault}, {Staveley-Smith}, {Subrahmanyan}, {Walker},
  \& {Wilson}}]{at20g}
{Murphy}, T., {Sadler}, E.~M., {Ekers}, R.~D., {et~al.} 2010, \mnras, 402, 2403

\bibitem[{Peloton(2015)}]{peloton:tel-01257383}
Peloton, J. 2015, Theses, {Universite Paris Diderot-Paris VII ; Laboratoire
  AstroParticule \& Cosmologie }

\bibitem[{{Planck Collaboration} {et~al.}(2014){Planck Collaboration}, {Ade},
  {Aghanim}, {Alves}, {Armitage-Caplan}, {Arnaud}, {Ashdown},
  {Atrio-Barandela}, {Aumont}, {Aussel}, \& et~al.}]{2014A&A...571A...1P}
{Planck Collaboration}, {Ade}, P.~A.~R., {Aghanim}, N., {et~al.} 2014, \aap,
  571, A1

\bibitem[{{Planck Collaboration} {et~al.}(2016{\natexlab{a}}){Planck
  Collaboration}, {Adam}, {Ade}, {Aghanim}, {Akrami}, {Alves}, {Arg{\"u}eso},
  {Arnaud}, {Arroja}, {Ashdown}, \& et~al.}]{2016A&A...594A...1P}
{Planck Collaboration}, {Adam}, R., {Ade}, P.~A.~R., {et~al.}
  2016{\natexlab{a}}, \aap, 594, A1

\bibitem[{{Planck Collaboration} {et~al.}(2016{\natexlab{b}}){Planck
  Collaboration}, {Adam}, {Ade}, {Aghanim}, {Arnaud}, {Ashdown}, {Aumont},
  {Baccigalupi}, {Banday}, {Barreiro}, \& et~al.}]{2016A&A...594A...9P}
---. 2016{\natexlab{b}}, \aap, 594, A9

\bibitem[{{Planck Collaboration} {et~al.}(2016{\natexlab{c}}){Planck
  Collaboration}, {Adam}, {Ade}, {Aghanim}, {Alves}, {Arnaud}, {Ashdown},
  {Aumont}, {Baccigalupi}, {Banday}, \& et~al.}]{2016A&A...594A..10P}
---. 2016{\natexlab{c}}, \aap, 594, A10

\bibitem[{{Planck Collaboration} {et~al.}(2016{\natexlab{d}}){Planck
  Collaboration}, {Ade}, {Aghanim}, {Arnaud}, {Arroja}, {Ashdown}, {Aumont},
  {Baccigalupi}, {Ballardini}, {Banday}, \& et~al.}]{planck2015XX}
{Planck Collaboration}, {Ade}, P.~A.~R., {Aghanim}, N., {et~al.}
  2016{\natexlab{d}}, \aap, 594, A20

\bibitem[{{Planck Collaboration} {et~al.}(2016{\natexlab{e}}){Planck
  Collaboration}, {Akrami}, {Ashdown}, {Aumont}, {Baccigalupi}, {Ballardini},
  {Banday}, {Barreiro}, {Bartolo}, {Basak}, {Benabed}, {Bernard}, {Bersanelli},
  {Bielewicz}, {Bonavera}, {Bond}, {Borrill}, {Bouchet}, {Boulanger}, {Bucher},
  {Burigana}, {Butler}, {Calabrese}, {Cardoso}, {Carron}, {Chiang}, {Colombo},
  {Comis}, {Couchot}, {Coulais}, {Crill}, {Curto}, {Cuttaia}, {de Bernardis},
  {de Rosa}, {de Zotti}, {Delabrouille}, {Di Valentino}, {Dickinson}, {Diego},
  {Dor{\'e}}, {Ducout}, {Dupac}, {Elsner}, {En{\ss}lin}, {Eriksen},
  {Falgarone}, {Fantaye}, {Finelli}, {Frailis}, {Fraisse}, {Franceschi},
  {Frolov}, {Galeotta}, {Galli}, {Ganga}, {G{\'e}nova-Santos}, {Gerbino},
  {Gonz{\'a}lez-Nuevo}, {G{\'o}rski}, {Gruppuso}, {Gudmundsson}, {Hansen},
  {Helou}, {Henrot-Versill{\'e}}, {Herranz}, {Hivon}, {Jaffe}, {Jones},
  {Keih{\"a}nen}, {Keskitalo}, {Kiiveri}, {Kim}, {Kisner}, {Krachmalnicoff},
  {Kunz}, {Kurki-Suonio}, {Lagache}, {Lamarre}, {Lasenby}, {Lattanzi},
  {Lawrence}, {Le Jeune}, {Lellouch}, {Levrier}, {Liguori}, {Lilje},
  {Lindholm}, {L{\'o}pez-Caniego}, {Ma}, {Mac{\'{\i}}as-P{\'e}rez}, {Maggio},
  {Maino}, {Mandolesi}, {Maris}, {Martin}, {Mart{\'{\i}}nez-Gonz{\'a}lez},
  {Matarrese}, {Mauri}, {McEwen}, {Melchiorri}, {Mennella}, {Migliaccio},
  {Miville-Desch{\^e}nes}, {Molinari}, {Moneti}, {Montier}, {Moreno},
  {Morgante}, {Natoli}, {Oxborrow}, {Paoletti}, {Partridge}, {Patanchon},
  {Patrizii}, {Perdereau}, {Piacentini}, {Plaszczynski}, {Polenta}, {Rachen},
  {Racine}, {Reinecke}, {Remazeilles}, {Renzi}, {Rocha}, {Romelli}, {Rosset},
  {Roudier}, {Rubi{\~n}o-Mart{\'{\i}}n}, {Ruiz-Granados}, {Salvati}, {Sandri},
  {Savelainen}, {Scott}, {Sirri}, {Spencer}, {Suur-Uski}, {Tauber},
  {Tavagnacco}, {Tenti}, {Toffolatti}, {Tomasi}, {Tristram}, {Trombetti},
  {Valiviita}, {Van Tent}, {Vielva}, {Villa}, {Wehus}, \&
  {Zacchei}}]{2016arXiv161207151P}
{Planck Collaboration}, {Akrami}, Y., {Ashdown}, M., {et~al.}
  2016{\natexlab{e}}, ArXiv e-prints, arXiv:1612.07151

\bibitem[{{Planck Collaboration Int. XXII}(2015)}]{planck2014-XXII}
{Planck Collaboration Int. XXII}. 2015, A\&A, 576, A107

\bibitem[{{Planck Collaboration Int. XXX}(2016)}]{2016A&A...586A.133P}
{Planck Collaboration Int. XXX}. 2016, A\&A, 586, A133

\bibitem[{{Poletti} {et~al.}(2016){Poletti}, {Fabbian}, {Le Jeune}, {Peloton},
  {Arnold}, {Baccigalupi}, {Barron}, {Beckman}, {Borrill}, {Chapman},
  {Chinone}, {Cukierman}, {Ducout}, {Elleflot}, {Errard}, {Feeney},
  {Goeckner-Wald}, {Groh}, {Hall}, {Hasegawa}, {Hazumi}, {Hill}, {Howe},
  {Inoue}, {Jaffe}, {Jeong}, {Katayama}, {Keating}, {Keskitalo}, {Kisner},
  {Kusaka}, {Lee}, {Leon}, {Linder}, {Lowry}, {Matsuda}, {Navaroli}, {Paar},
  {Puglisi}, {Reichardt}, {Ross}, {Siritanasak}, {Stebor}, {Steinbach},
  {Stompor}, {Suzuki}, {Tajima}, {Teply}, \& {Whitehorn}}]{explicit_map_making}
{Poletti}, D., {Fabbian}, G., {Le Jeune}, M., {et~al.} 2016, \aap, 600, A60

\bibitem[{{QUIET Collaboration} {et~al.}(2012){QUIET Collaboration}, {Araujo},
  {Bischoff}, {Brizius}, {Buder}, {Chinone}, {Cleary}, {Dumoulin}, {Kusaka},
  {Monsalve}, {N{\ae}ss}, {Newburgh}, {Reeves}, {Wehus}, {Zwart}, {Bronfman},
  {Bustos}, {Church}, {Dickinson}, {Eriksen}, {Gaier}, {Gundersen}, {Hasegawa},
  {Hazumi}, {Huffenberger}, {Ishidoshiro}, {Jones}, {Kangaslahti}, {Kapner},
  {Kubik}, {Lawrence}, {Limon}, {McMahon}, {Miller}, {Nagai}, {Nguyen},
  {Nixon}, {Pearson}, {Piccirillo}, {Radford}, {Readhead}, {Richards},
  {Samtleben}, {Seiffert}, {Shepherd}, {Smith}, {Staggs}, {Tajima}, {Thompson},
  {Vanderlinde}, \& {Williamson}}]{2012ApJ...760..145Q}
{QUIET Collaboration}, {Araujo}, D., {Bischoff}, C., {et~al.} 2012, \apj, 760,
  145

\bibitem[{{Smith}(2006)}]{Smith:2006}
{Smith}, K.~M. 2006, \prd, 74, 083002

\bibitem[{{Suzuki} {et~al.}(2016){Suzuki}, {Ade}, {Akiba}, {Aleman}, {Arnold},
  {Baccigalupi}, {Barch}, {Barron}, {Bender}, {Boettger}, {Borrill}, {Chapman},
  {Chinone}, {Cukierman}, {Dobbs}, {Ducout}, {Dunner}, {Elleflot}, {Errard},
  {Fabbian}, {Feeney}, {Feng}, {Fujino}, {Fuller}, {Gilbert}, {Goeckner-Wald},
  {Groh}, {Haan}, {Hall}, {Halverson}, {Hamada}, {Hasegawa}, {Hattori},
  {Hazumi}, {Hill}, {Holzapfel}, {Hori}, {Howe}, {Inoue}, {Irie}, {Jaehnig},
  {Jaffe}, {Jeong}, {Katayama}, {Kaufman}, {Kazemzadeh}, {Keating}, {Kermish},
  {Keskitalo}, {Kisner}, {Kusaka}, {Jeune}, {Lee}, {Leon}, {Linder}, {Lowry},
  {Matsuda}, {Matsumura}, {Miller}, {Mizukami}, {Montgomery}, {Navaroli},
  {Nishino}, {Peloton}, {Poletti}, {Puglisi}, {Rebeiz}, {Raum}, {Reichardt},
  {Richards}, {Ross}, {Rotermund}, {Segawa}, {Sherwin}, {Shirley},
  {Siritanasak}, {Stebor}, {Stompor}, {Suzuki}, {Tajima}, {Takada}, {Takakura},
  {Takatori}, {Tikhomirov}, {Tomaru}, {Westbrook}, {Whitehorn}, {Yamashita},
  {Zahn}, \& {Zahn}}]{2016JLTP..184..805S}
{Suzuki}, A., {Ade}, P., {Akiba}, Y., {et~al.} 2016, Journal of Low Temperature
  Physics, 184, 805

\bibitem[{Takakura {et~al.}(2017)Takakura, Aguilar, Akiba, Arnold, Baccigalupi,
  Barron, Beckman, Boettger, Borrill, Chapman, Chinone, Cukierman, Ducout,
  Elleflot, Errard, Fabbian, Fujino, Galitzki, Goeckner-Wald, Halverson,
  Hasegawa, Hattori, Hazumi, Hill, Howe, Inoue, Jaffe, Jeong, Kaneko, Katayama,
  Keating, Keskitalo, Kisner, Krachmalnicoff, Kusaka, Lee, Leon, Lowry,
  Matsuda, Matsumura, Navaroli, Nishino, Paar, Peloton, Poletti, Puglisi,
  Reichardt, Ross, Siritanasak, Suzuki, Tajima, Takatori, \&
  Teply}]{1475-7516-2017-05-008}
Takakura, S., Aguilar, M., Akiba, Y., {et~al.} 2017, Journal of Cosmology and
  Astroparticle Physics, 2017, 008

\bibitem[{{The \textsc{Bicep2} and Keck Array Collaborations: P.~A.~R.~Ade}
  {et~al.}(2015){The \textsc{Bicep2} and Keck Array Collaborations:
  P.~A.~R.~Ade}, {Ahmed}, {Aikin}, {Alexander}, {Barkats}, {Benton},
  {Bischoff}, {Bock}, {Brevik}, {Buder}, {Bullock}, {Buza}, {Connors}, {Crill},
  {Dowell}, {Dvorkin}, {Duband}, {Filippini}, {Fliescher}, {Golwala},
  {Halpern}, {Harrison}, {Hasselfield}, {Hildebrandt}, {Hilton}, {Hristov},
  {Hui}, {Irwin}, {Karkare}, {Kaufman}, {Keating}, {Kefeli}, {Kernasovskiy},
  {Kovac}, {Kuo}, {Leitch}, {Lueker}, {Mason}, {Megerian}, {Netterfield},
  {Nguyen}, {O'Brient}, {Ogburn}, {Orlando}, {Pryke}, {Reintsema}, {Richter},
  {Schwarz}, {Sheehy}, {Staniszewski}, {Sudiwala}, {Teply}, {Thompson},
  {Tolan}, {Turner}, {Vieregg}, {Weber}, {Willmert}, {Wong}, \& {Yoon}}]{bkV}
{The \textsc{Bicep2} and Keck Array Collaborations: P.~A.~R.~Ade}, {Ahmed}, Z.,
  {Aikin}, R.~W., {et~al.} 2015, \apj, 811, 126

\bibitem[{{The \textsc{Bicep2} and Keck Array Collaborations: P.~A.~R.~Ade}
  {et~al.}(2016){The \textsc{Bicep2} and Keck Array Collaborations:
  P.~A.~R.~Ade}, {Ahmed}, {Aikin}, {Alexander}, {Barkats}, {Benton},
  {Bischoff}, {Bock}, {Bowens-Rubin}, {Brevik}, {Buder}, {Bullock}, {Buza},
  {Connors}, {Crill}, {Duband}, {Dvorkin}, {Filippini}, {Fliescher}, {Grayson},
  {Halpern}, {Harrison}, {Hilton}, {Hui}, {Irwin}, {Karkare}, {Karpel},
  {Kaufman}, {Keating}, {Kefeli}, {Kernasovskiy}, {Kovac}, {Kuo}, {Leitch},
  {Lueker}, {Megerian}, {Netterfield}, {Nguyen}, {O'Brient}, {Ogburn},
  {Orlando}, {Pryke}, {Richter}, {Schwarz}, {Sheehy}, {Staniszewski},
  {Steinbach}, {Sudiwala}, {Teply}, {Thompson}, {Tolan}, {Tucker}, {Turner},
  {Vieregg}, {Weber}, {Wiebe}, {Willmert}, {Wong}, {Wu}, \& {Yoon}}]{bkVI}
---. 2016, \prl, 116, 031302

\bibitem[{{The \textsc{Bicep2} Collaboration: P.~A.~R.~Ade} {et~al.}(2014){The
  \textsc{Bicep2} Collaboration: P.~A.~R.~Ade}, {Aikin}, {Barkats}, {Benton},
  {Bischoff}, {Bock}, {Brevik}, {Buder}, {Bullock}, {Dowell}, {Duband},
  {Filippini}, {Fliescher}, {Golwala}, {Halpern}, {Hasselfield}, {Hildebrandt},
  {Hilton}, {Hristov}, {Irwin}, {Karkare}, {Kaufman}, {Keating},
  {Kernasovskiy}, {Kovac}, {Kuo}, {Leitch}, {Lueker}, {Mason}, {Netterfield},
  {Nguyen}, {O'Brient}, {Ogburn}, {Orlando}, {Pryke}, {Reintsema}, {Richter},
  {Schwarz}, {Sheehy}, {Staniszewski}, {Sudiwala}, {Teply}, {Tolan}, {Turner},
  {Vieregg}, {Wong}, \& {Yoon}}]{bkI}
{The \textsc{Bicep2} Collaboration: P.~A.~R.~Ade}, {Aikin}, R.~W., {Barkats},
  D., {et~al.} 2014, \prl, 112, 241101

\bibitem[{{The \textsc{Bicep2}/Keck and Planck Collaborations: P.~A.~R.~Ade}
  {et~al.}(2015){The \textsc{Bicep2}/Keck and Planck Collaborations:
  P.~A.~R.~Ade}, {Aghanim}, {Ahmed}, {Aikin}, {Alexander}, {Arnaud}, {Aumont},
  {Baccigalupi}, {Banday}, \& et~al.}]{bkp}
{The \textsc{Bicep2}/Keck and Planck Collaborations: P.~A.~R.~Ade}, {Aghanim},
  N., {Ahmed}, Z., {et~al.} 2015, \prl, 114, 101301

\bibitem[{{The \textsc{Polarbear} Collaboration: P.~A.~R.~Ade}
  {et~al.}(2014{\natexlab{a}}){The \textsc{Polarbear} Collaboration:
  P.~A.~R.~Ade}, {Akiba}, {Anthony}, {Arnold}, {Atlas}, {Barron}, {Boettger},
  {Borrill}, {Chapman}, {Chinone}, {Dobbs}, {Elleflot}, {Errard}, {Fabbian},
  {Feng}, {Flanigan}, {Gilbert}, {Grainger}, {Halverson}, {Hasegawa},
  {Hattori}, {Hazumi}, {Holzapfel}, {Hori}, {Howard}, {Hyland}, {Inoue},
  {Jaehnig}, {Jaffe}, {Keating}, {Kermish}, {Keskitalo}, {Kisner}, {Le Jeune},
  {Lee}, {Leitch}, {Linder}, {Lungu}, {Matsuda}, {Matsumura}, {Meng}, {Miller},
  {Morii}, {Moyerman}, {Myers}, {Navaroli}, {Nishino}, {Orlando}, {Paar},
  {Peloton}, {Poletti}, {Quealy}, {Rebeiz}, {Reichardt}, {Richards}, {Ross},
  {Schanning}, {Schenck}, {Sherwin}, {Shimizu}, {Shimmin}, {Shimon},
  {Siritanasak}, {Smecher}, {Spieler}, {Stebor}, {Steinbach}, {Stompor},
  {Suzuki}, {Takakura}, {Tomaru}, {Wilson}, {Yadav}, \& {Zahn}}]{pb2014c}
{The \textsc{Polarbear} Collaboration: P.~A.~R.~Ade}, {Akiba}, Y., {Anthony},
  A.~E., {et~al.} 2014{\natexlab{a}}, \apj, 794, 171

\bibitem[{{The \textsc{Polarbear} Collaboration: P.~A.~R.~Ade}
  {et~al.}(2014{\natexlab{b}}){The \textsc{Polarbear} Collaboration:
  P.~A.~R.~Ade}, {Akiba}, {Anthony}, {Arnold}, {Atlas}, {Barron}, {Boettger},
  {Borrill}, {Chapman}, {Chinone}, {Dobbs}, {Elleflot}, {Errard}, {Fabbian},
  {Feng}, {Flanigan}, {Gilbert}, {Grainger}, {Halverson}, {Hasegawa},
  {Hattori}, {Hazumi}, {Holzapfel}, {Hori}, {Howard}, {Hyland}, {Inoue},
  {Jaehnig}, {Jaffe}, {Keating}, {Kermish}, {Keskitalo}, {Kisner}, {Le Jeune},
  {Lee}, {Linder}, {Leitch}, {Lungu}, {Matsuda}, {Matsumura}, {Meng}, {Miller},
  {Morii}, {Moyerman}, {Myers}, {Navaroli}, {Nishino}, {Paar}, {Peloton},
  {Quealy}, {Rebeiz}, {Reichardt}, {Richards}, {Ross}, {Schanning}, {Schenck},
  {Sherwin}, {Shimizu}, {Shimmin}, {Shimon}, {Siritanasak}, {Smecher},
  {Spieler}, {Stebor}, {Steinbach}, {Stompor}, {Suzuki}, {Takakura}, {Tomaru},
  {Wilson}, {Yadav}, \& {Zahn}}]{pb2014b}
---. 2014{\natexlab{b}}, \prl, 113, 021301

\bibitem[{Weiland {et~al.}(2011)Weiland, Odegard, Hill, Wollack, Hinshaw,
  Greason, Jarosik, Page, Bennett, Dunkley, {et~al.}}]{weiland2011seven}
Weiland, J., Odegard, N., Hill, R., {et~al.} 2011, The Astrophysical Journal
  Supplement Series, 192, 19

\bibitem[{Wrobel {et~al.}(1998)Wrobel, Patnaik, Browne, \&
  Wilkinson}]{vlanorth}
Wrobel, J.~M., Patnaik, A.~R., Browne, I.~W.~A., \& Wilkinson, P.~N. 1998,
  \baas, 30, 1308

\end{thebibliography}

\end{document}